\documentclass[12pt]{article}
\usepackage{amsfonts}
\usepackage{amssymb}
\usepackage{amsbsy}
\usepackage{graphicx}    
\usepackage{rotating}    
\usepackage{epsfig}
\usepackage{color}

\input epsf.sty

\newlength{\TZ}
\newcommand{\BEQ}{\begin{equation}}     
\newcommand{\BEA}{\begin{eqnarray}}
\newcommand{\BD}{\begin{displaymath}}
\newcommand{\EEQ}{\end{equation}}       
\newcommand{\EEA}{\end{eqnarray}}
\newcommand{\ED}{\end{displaymath}}
\newcommand{\eps}{\varepsilon}          
\newcommand{\vph}{\varphi}              
\newcommand{\D}{{\rm d}}                
\newcommand{\II}{{\rm i}}               
\newcommand{\demi}{\frac{1}{2}}         
\newcommand{\wit}[1]{\widetilde{#1}}    
\newcommand{\wht}[1]{\widehat{#1}}      

\renewcommand{\vec}[1]{\boldsymbol{#1}} 

\catcode`\@=11
\def\numberbysection{\@addtoreset{equation}{section}
        \def\theequation{\thesection.\arabic{equation}}}
\numberbysection

\begin{document}

\begin{titlepage}

\vskip 1.5 cm
\begin{center}
{\Large \bf Dynamical symmetries and causality \\[0.1truecm] in non-equilibrium phase transitions}
\end{center}

\vskip 2.0 cm
\centerline{{\bf Malte Henkel}} 
\vskip 0.5 cm
\begin{center}
Groupe de Physique Statistique,
D\'epartement de Physique de la Mati\`ere et des Mat\'eriaux,
Institut Jean Lamour (CNRS UMR 7198), Universit\'e de Lorraine Nancy, 
B.P. 70239, \\ F -- 54506 Vand{\oe}uvre l\`es Nancy Cedex, France\\
\end{center}

\begin{abstract}
Dynamical symmetries are of considerable importance in elucidating the complex behaviour of strongly 
interacting systems with many degrees of freedom.
Paradigmatic examples are cooperative phenomena as they arise in phase transitions,
where conformal invariance has led to enormous progress in equilibrium phase transitions, especially in two dimensions. 
Non-equilibrium phase transitions can arise in much larger portions of the parameter space than equilibrium phase transitions. 
The state of the art of recent attempts to generalise conformal invariance
to a new generic symmetry, taking into account the different scaling behaviour of space and time, 
will be reviewed. Particular attention will be given to the causality
properties as they follow for co-variant $n$-point functions. 
These are important for the physical identification of $n$-point functions as responses or correlators.
\end{abstract}

\vfill

\end{titlepage}

\setcounter{footnote}{0} 


\section{Introduction}

Improving our understanding of the collective behaviour of 
strongly interacting systems consisting of a large number of strongly interacting
degrees of freedom is an ongoing challenge. From the point of view of 
the statistical physicist, paradigmatic examples are provided  by
systems undergoing a continuous phase transition, where fluctuation 
effects render traditional methods such as mean-field approximations inapplicable \cite{Henkel10,Taeuber14}. 
At the same time, it turns out that
these systems can be effectively characterised in terms of a small number of `relevant' scaling operators, such that the
net effect of all other physical quantities, the `irrelevant' ones, merely amounts to the generation of corrections to the leading
scaling behaviour. From a symmetry perspective, phase transitions naturally acquire some kind of scale-invariance and it then
becomes a natural question whether further dynamical symmetries can be present. 

{\bf 1.} In {\em equilibrium critical phenomena} (for systems with sufficiently short-ranged, 
local interactions) scale-invariance is extended to conformal
invariance. In two space dimensions, the generators $\ell_n, \bar{\ell}_n$ should obey the infinite-dimensional algebra
\BEQ \label{1}
\left[ \ell_n, \ell_m \right] = (n-m) \ell_{n+m}, \quad \left[ \bar{\ell}_n, \bar{\ell}_m\right] = (n-m)\bar{\ell}_{n+m},\quad
\left[ \ell_n, \bar{\ell}_m\right] =0
\EEQ
for $n,m\in\mathbb{Z}$. 
The action of these generators on physical scaling operators $\phi(z,\bar{z})$, 
where complex coordinates $z,\bar{z}$ are used, 
is conventionally given by the representation \cite{Cartan1909,Belavin84}
\BEQ \label{2}
\ell_n \phi(z,\bar{z}) \to \left[\ell_n, \phi(z,\bar{z})\right]
 = -\left( z^{n+1} \partial_z +\Delta (n+1) z^n \right)\phi(z,\bar{z})
\EEQ
and similarly for $\bar{\ell}_n$, where the r\^oles of $z$ and $\bar{z}$ are exchanged. Herein, the {\em conformal weights} $\Delta,\overline{\Delta}$ are real
constants, and related to the scaling dimension $x_{\phi}=\Delta+\overline{\Delta}$ 
and the spin $s_{\phi}=\Delta-\overline{\Delta}$ of the
scaling operator $\phi$. The representation (\ref{2}) is an infinitesimal form of the (anti)holomorphic transformations 
$z\mapsto w(z)$ and 
$\bar{z}\mapsto \bar{w}(\bar{z})$. The maximal finite-dimensional sub-algebra of (\ref{1}) is isomorphic to 
$\mathfrak{sl}(2,\mathbb{R})\oplus\mathfrak{sl}(2,\mathbb{R})\cong\left\langle \ell_{\pm 1,0}, \bar{\ell}_{\pm 1,0}\right\rangle$. 
It is this conformal
sub-algebra only which has an analogue in higher space dimensions $d>2$. 
Denoting the Laplace operator by ${\cal S}:=4\partial_z \partial_{\bar{z}}=4\ell_{-1}\bar{\ell}_{-1}$, 
the conformal invariance of the Laplace equation ${\cal S}\phi(z,\bar{z})=0$ is expressed through the commutator
\BEQ
\left[ {\cal S}, \ell_n \right]\phi(z,\bar{z}) = -(n+1) z^n {\cal S}\phi(z,\bar{z}) -4\Delta(n+1)n z^{n-1}\partial_{\bar{z}} \phi(z,\bar{z})
\EEQ
and analogously for $\bar{\ell}_n$. Hence, for vanishing conformal weights $\Delta=\Delta_{\phi}=0$ and $\overline{\Delta}=\overline{\Delta}_{\phi}=0$, 
any solution of ${\cal S}\phi=0$ is mapped onto another solution of the same equation. 
Beyond, the Laplace equation, thermal fluctuations in $2D$ classical critical points or
quantum fluctuations in $1D$ quantum critical points (at temperature $T=0$) 
modify the conformal algebra (\ref{1}) to a pair of commuting
Virasoro algebras, parametrised by the central charge $c$. Then eq.~(\ref{2}) 
retains its validity when the set of admissible operators $\phi$ is
restricted to the set of {\em primary scaling operators} (a scaling operator is called {\em quasi-primary} if the transformation (\ref{2}) only holds for
the finite-dimensional sub-algebra $\mathfrak{sl}(2,\mathbb{R})\cong\left\langle \ell_{\pm 1,0}\right\rangle$) \cite{Belavin84}. 
In turn, this furnishes the basis for the derivation of {\em conformal Ward identities} obeyed by $n$-point correlation functions 
$F_n := \left\langle \phi_1(z_1,\bar{z}_1)\ldots\phi_n(z_n,\bar{z}_n)\right\rangle$ of primary operators $\phi_1\ldots \phi_n$. 
Celebrate theorems provide a classification of the Virasoro primary operators from the unitary 
representations of the Virasoro algebra, for example through the Kac formula for central charges $c<1$ \cite{Francesco97,Unterberger11}. 
Numerous physical applications continue being discovered.  

{\bf 2.} When turning to time-dependent critical phenomena, the theory is far less advanced. One of the best-studied examples is the
{\em Schr\"odinger-Virasoro algebra} $\mathfrak{sv}(d)$ in $d$ space dimensions \cite{Henkel94,Henkel02}
\BEA
{} \bigl[ X_n, X_{n'}\bigr] &=& (n-n') X_{n+n'} \hspace{1.4truecm} \;\; , \;\; \hspace{0.33truecm}
{} \bigl[ X_n, Y_m^{(j)}\bigr] \:=\: \left(\frac{n}{2} -m\right) Y_{n+m}^{(j)}
\nonumber \\
{} \bigl[ X_n, M_{n'} \bigr] &=& - n' M_{n+n'} \hspace{2.1truecm}\;\; , \;\; \hspace{0.33truecm}
{} \bigl[ X_n, R_{n'}^{(jk)} \bigr] \:=\: -n' R_{n+n'}^{(jk)} \label{4} \\
{} \bigl[ Y_{m}^{(j)}, Y_{m'}^{(k)} \bigr] &=& \delta^{j,k}\,
\left (m - m'\right) M_{m+m'}  \hspace{0.2truecm} \;\; , \;\; 
\hspace{0.33truecm} \bigl[ R_n^{(jk)},Y_m^{(\ell)} \bigr] 
\:=\: \delta^{j,\ell}\, Y_{n+m}^{(k)} -
\delta^{k,\ell}\, Y_{n+m}^{(j)}  \nonumber \\
{} \bigl[ R_n^{(jk)},R_{n'}^{(\ell i)} \bigr] &=& 
\delta^{j,i}R_{n+n'}^{(\ell k)} - \delta^{k,\ell}R_{n+n'}^{(ji)} + \delta^{k,i}R_{n+n'}^{(j\ell)} - \delta^{j,\ell}R_{n+n'}^{(ik)}
\nonumber
\EEA 
(all other commutators vanish) with integer indices $n,n'\in\mathbb{Z}$,  half-integer indices $m,m'\in\mathbb{Z}+\demi$ 
and $i,j,k,\ell\in\{1,\ldots,d\}$. Casting the generators of $\mathfrak{sv}(d)$ into the four families
$X,Y^{(j)}, M, R^{(jk)}=-R^{(kj)}$ makes explicit (i) that the generators $X_n$ 
form a conformal sub-algebra and (ii) that the families $Y^{(j)}$ and
$M, R^{(jk)}$ make up Virasoro primary operators of weight $\frac{3}{2}$ and $1$, respectively \cite{Henkel94}. 
Non-trivial central extensions are only possible (i) either in the
conformal sub-algebra $\left\langle X_n\right\rangle_{n\in\mathbb{Z}}$, 
where it must be of the form of the Virasoro central charge, or else (ii) in the
$\mathfrak{so}(d)$-current algebra $\left\langle R_{n}^{(jk)}\right\rangle_{n\in\mathbb{Z}}$, 
where it must be a Kac-Moody central charge \cite{Henkel94,Francesco97,Roger06,Unterberger11}. 
The maximal finite-dimensional sub-algebra of
$\mathfrak{sv}(d)$ is the {\em Schr\"odinger algebra} 
$\mathfrak{sch}(d) = \left\langle X_{0,\pm 1}, Y_{\pm 1/2}^{(j)}, M_0, R_0^{(jk)}\right\rangle_{j,k=1,\ldots d}$, where $M_0$ is central.  
An explicit representation in terms of time-space coordinates 
$(t,\vec{r})\in\mathbb{R}\times\mathbb{R}^d$, acting on a (scalar) scaling operator 
$\phi(t,\vec{r})$ of scaling dimension $x$ and of mass $\cal M$, is given by \cite{Henkel94}
\BEA
X_n &=& -t^{n+1}\partial_t - \frac{n+1}{2}t^n \vec{r}\cdot\vec{\nabla}_{\vec{r}} 
- \frac{\cal M}{4}(n+1)n t^{n-1} \vec{r}^2
- \frac{n+1}{2} x t^n \nonumber \\
Y_m^{(j)} &=& - t^{m+1/2} \partial_j - \left( m + \demi\right)  t^{m-1/2} {\cal M} r_j \nonumber \\
M_n &=& - t^n {\cal M}  \label{5} \\
R_n^{(jk)} &=& - t^n \bigl( r_j \partial_k - r_k \partial_j \bigr) \:=\: - R_n^{(kj)} \nonumber
\EEA
with the abbreviations $\partial_j := \partial/\partial r_j$ and $\vec{\nabla}_{\vec{r}} = (\partial_1, \ldots, \partial_d)^{\rm T}$. 
These are the infinitesimal forms of the transformations $(t,\vec{r})\mapsto (t',\vec{r}')$, where
\BEA
X_n:&& t =\beta(t'), \quad \vec{r}=\vec{r}' \sqrt{\frac{\D\beta(t')}{\D t'}} \nonumber \\
Y_m:&& t = t',\quad \hspace{0.6truecm}\vec{r}=\vec{r}'-\alpha(t') \\
R_n:&& t = t',\quad \hspace{0.6truecm}\vec{r} = {\cal R}(t')\vec{r}' \nonumber
\EEA
where $\alpha(t)$ is an arbitrary time-dependent function, $\beta(t)$ is a non-decreasing function 
and ${\cal R}(t)\in{\sl SO}(d)$ denotes a rotation matrix with time-dependent rotation angles. The generators $M_n$ do not generate
a time-space transformation, but rather produce a time-dependent `phase shift' of the scaling operator $\phi$.\footnote{To see this explicitly,  
one should exponentiate these generators to create their corresponding finite transformations, see \cite{Henkel03a}.} 

The dilatations $X_0$ are the infinitesimal form of the transformations 
$t\mapsto \lambda^z t$ and $\vec{r}\mapsto \lambda \vec{r}$, where
$\lambda\in\mathbb{R}_+$ is a constant and $z$ is called the {\em dynamical exponent}. 
In the representation (\ref{5}), one has $z=2$. 

Since the work of Lie \cite{Lie1881}, and before of Jacobi \cite{Jacobi1843}, 
the Schr\"odinger algebra is known to be a dynamic symmetry of the free diffusion equation (and, much later, also of the free Schr\"odinger equation). 
Define the Schr\"odinger operator
\BEQ
{\cal S} = 2{\cal M}\partial_t - \vec{\nabla}_{\vec{r}}\cdot\vec{\nabla}_{\vec{r}} = 2 M_0 X_{-1} - \vec{Y}_{-1/2}\cdot\vec{Y}_{-1/2}
\EEQ
Following Niederer \cite{Niederer72}, dynamical symmetries of such linear equations are analysed through the commutators of
$\cal S$ with the symmetry Lie algebra. For the case of $\mathfrak{sch}(d)$, the only non-vanishing commutators with $\cal S$ are
\BEQ \label{8}
\left[ {\cal S},X_0\right] = -{\cal S},\quad \left[{\cal S},X_{1}\right] = -2t{\cal S} -(2x-d)M_0
\EEQ
Hence any solution $\phi$ of the free Schr\"odinger/diffusion equation
${\cal S}\phi=0$ with scaling dimension $x_{\phi}=\frac{d}{2}$ is mapped onto another solution of 
the free Schr\"odinger equation.\footnote{Unitarity of the representation implies the bound $x\geq \frac{d}{2}$ \cite{Lee09}.} 
Finally, from representations such as (\ref{5}), one can derive Schr\"odinger-Ward identities in order
to derive the form of covariant two-point functions 
$\left\langle \phi_1(t_1,\vec{r}_1)\ldots\phi_n(t_n,\vec{r}_n)\right\rangle$. With respect
to conformal invariance, one has the important difference that the generator 
$M_0=-{\cal M}$ is central in the finite-dimensional sub-algebra
$\mathfrak{sch}(d)$. This implies the {\em Bargman superselection rule} \cite{Bargman54} 
\BEQ \label{9} 
\left( {\cal M}_1 + \cdots + {\cal M}_n \right)\: \left\langle\phi_1(t_1,\vec{r}_1)\ldots\phi_n(t_n,\vec{r}_n)\right\rangle =0
\EEQ
Physicists' conventions require that `physical masses' ${\cal M}_i\geq 0$. It it therefore necessary to define a formal `complex conjugate'
$\phi^*$ of the scaling operator $\phi$, such that its mass ${\cal M}^* := - {\cal M}\leq 0$ 
becomes negative. Then one may write e.g. a non-vanishing
co-variant two-point function of two quasi-primary scaling operators 
(up to an undetermined constant of normalisation) \cite{Henkel94}
\BEQ \label{10}
\left\langle \phi_1(t_1,\vec{r}_1) \phi_2^*(t_2,\vec{r}_2)\right\rangle = \delta_{x_1,x_2}\delta({\cal M}_1-{\cal M}_2^*)\, 
(t_1-t_2)^{-x_1} \exp\left[-\frac{{\cal M}_1}{2} \frac{ (\vec{r}_1-\vec{r}_2)^2}{t_1-t_2} \right]
\EEQ
Here, and throughout this paper, $\delta_{a,b}=1$ if $a=b$ and $\delta_{a,b}=0$ if $a\ne b$. 
While eq.~(\ref{10}) looks at first sight like a reasonable heat kernel, a closer inspection raises several questions:
\begin{enumerate}
\item why should it be obvious that the time difference $t_1-t_2>0$, to make the power-law prefactor real-valued~? 
\item given the convention that ${\cal M}_1\geq 0$, the condition $t_1-t_2>0$ is also required in order to have a decay of
the $n$-point function with increasing distance $|\vec{r}|=|\vec{r}_1-\vec{r}_2|\to\infty$. 
\item in applications to non-equilibrium statistical physics, one studies indeed two-point functions of the above type, which
are then interpreted as the linear response function of the scaling operator 
$\phi$ with respect to an external conjugate field $h(t,\vec{r})$
\BEQ
R(t_1,t_2;\vec{r}_1,\vec{r}_2) = \left. \frac{\delta\langle \phi(t_1,\vec{r}_1)\rangle}{\delta h(t_2,\vec{r}_2)}\right|_{h=0} =
\left\langle \phi(t_1,\vec{r}_1)\wit{\phi}(t_2,\vec{r}_2)\right\rangle
\EEQ
which in the context of the non-equilibrium Janssen-de Dominicis theory \cite{Taeuber14} 
can be re-expressed as a two-point function involving the
scaling operator $\phi$ and its associate {\em response operator} $\wit{\phi}$. In this physical context, one has a natural 
interpretation of the `complex conjugate' in terms of the relationship of $\phi$ and $\wit{\phi}$.

Then, the formal condition $t_1-t_2>0$ simply becomes the {\em causality condition}, 
namely that a response will only arise at a later time
$t_1>t_2$ after the stimulation at time $t_2\geq 0$. 
\end{enumerate}
Hence it is necessary to inquire under what conditions the causality of Schr\"odinger-covariant 
$n$-point functions can be guaranteed.

{\bf 3.} Textbooks in quantum mechanics show that the Schr\"odinger equation is the non-relativistic variant of relativistic wave equations,
be it the Klein-Gordon equation for scalars or the Dirac equations for spinors. One might therefore expect that the Schr\"odinger algebra
could be obtained by a contraction from the conformal algebra, but this is untrue 
(although there is a well-known contraction from the Poincar\'e algebra to the Galilei sub-algebra). 
Rather, applying a contraction to the conformal algebra, one arrives at a 
different Lie algebra, which we call the {\em altern-Virasoro algebra}\footnote{The name was originally given since at that time, 
relationships with physical ageing ({\it altern} in German) were still expected.} \cite{Henkel97,Ovsienko98,Henkel03a} 
$\mathfrak{av}(d)=\left\langle X_n,Y_n^{(j)},R_n^{(jk)}\right\rangle_{n\in\mathbb{Z}}$ with 
$j,k=1,\ldots,d$, which nowadays is often referred to as {\em infinite conformal Galilean algebra}. 
Its non-vanishing commutators can be given as follows  
\BEA 
{} [X_n, X_{n'}] &=& (n-n') X_{n+n'} \;\;,\;\; 
{} [X_n, Y_{m}^{(j)}] \:=\: \left(n-m\right)Y_{n+m}^{(j)} \nonumber \\
{} [X_n, R_{n'}^{(jk)}] &=& -n' R_{n+n'}^{(jk)} \hspace{0.585truecm}\;\; , \;\; 
{} [R_n^{(jk)}, Y_m^{(\ell)}] 
\:=\: \delta^{j,\ell} Y_{n+m}^{(k)} - \delta^{k,\ell} Y_{n+m}^{(j)} \label{12} \\
{} [ R_n^{(jk)},R_{n'}^{(\ell i)}] &=& 
\delta^{j,i}R_{n+n'}^{(\ell k)} - \delta^{k,\ell}R_{n+n'}^{(ji)} + \delta^{k,i}R_{n+n'}^{(j\ell)} - \delta^{j,\ell}R_{n+n'}^{(ik)}
\nonumber
\EEA
An explicit representation as time-space transformation is \cite{Cherniha10}
\BEA 
X_n &=& - t^{n+1}\partial_t - (n+1) t^n \vec{r}\cdot\vec{\nabla}_{\vec{r}}
- n(n+1) t^{n-1} \vec{\gamma}\cdot\vec{r} - x (n+1)t^n
\nonumber \\
Y_n^{(j)} &=& - t^{n+1} \partial_{j} - (n+1) t^n \gamma_j  \label{13} \\
R_n^{(jk)} &=& - t^n \left( r_j \partial_{k} -  r_k \partial_{j} \right)
- t^n \left( \gamma_j \partial_{\gamma_k}-\gamma_k
\partial_{\gamma_j}\right) \:=\: - R_n^{(kj)} \nonumber 
\EEA
where $\vec{\gamma}=(\gamma_1,\ldots,\gamma_d)$ is a vector of dimensionful constants, called {\em rapidities}, and 
$x$ is again a scaling dimension. The dynamical exponent $z=1$. 
The maximal finite-dimensional sub-algebra of $\mathfrak{av}(d)$ is the {\em conformal Galilean algebra} 
$\mbox{\sc cga}(d)=\langle X_{\pm 1,0},Y_{\pm 1,0}^{(j)},R_0^{(jk)}\rangle_{j,k=1,\ldots,d}$ 
\cite{Havas78,Henkel97,Negro97,Negro97b,Henkel02,Henkel03a,Bagchi09,Martelli09,Setare11}.\footnote{In the context 
of asymptotically flat $3D$ gravity, 
an isomorphic Lie algebra is known as BMS algebra, $\mathfrak{bms}_3\cong \mbox{\sc cga}(1)$ 
\cite{Barnich07,Barnich13,Bagchi12,Bagchi13b}.}  

A more abstract characterisation of $\mathfrak{av}(1)$ can be given in terms of $\alpha$-densities 
${\cal F}_{\alpha}=\{ u(z)(\D z)^{\alpha}\}$, with the action
\BEQ
f(z){\D\over \D z} \left(u(z)(\D z)^{\alpha}\right)
=(fu'+\alpha f' u)(z) (\D z)^{\alpha}.
\EEQ

\noindent {\bf Lemma 1.} {\rm \cite{Henkel06b}} 
{\it One has the isomorphism, where $\ltimes$ denotes the semi-direct sum}
\BEQ
\mathfrak{av}(1) \cong {\mathrm{Vect}}(S^1)\ltimes {\cal F}_{-1}
\EEQ
{\it Clearly, it follows that $\mbox{\sc cga}(1)\cong\mathfrak{sl}(2,\mathbb{R})\ltimes{\cal F}_{-1}$.} \\[0.02truecm]

As before, the time-space representation (\ref{13}) can be used to derive conformal-galilean Ward identities. For example, the
$\mbox{\sc cga}(d)$-covariant two-point function takes the form
\BEQ
\left\langle \phi_1(t_1,\vec{r}_1)\phi_2(t_2,\vec{r}_2)\right\rangle = \delta_{x_1,x_2}\delta_{\vec{\gamma}_1,\vec{\gamma}_2}\,
(t_1-t_2)^{-2x_1} \exp\left[ - \frac{\vec{\gamma}_1\cdot(\vec{r}_1-\vec{r}_2)}{t_1-t_2} \right]
\EEQ
Again, at first sight this looks physically reasonable, but several questions must be raised:
\begin{enumerate}
\item why should one have $t_1-t_2>0$ for the time difference, as required to make the power-law prefactor real-valued~? 
\item even for a fixed vector $\vec{\gamma}_1$ of rapidities, and even if $t_1-t_2>0$ could be taken for granted, how
does one guarantee that the scalar product $\vec{\gamma}_1\cdot(\vec{r}_1-\vec{r}_2)>0$, such that the two-point function
decreases as $|\vec{r}|=|\vec{r}_1-\vec{r}_2|\to\infty$~? 
\end{enumerate}

The finite-dimensional $\mbox{\sc cga}(2)$ admits a 
so-called `exotic' central extension \cite{Lukierski06,Lukierski07}. Abstractly, this is achieved
by completing the commutators (\ref{12}) by the following 
\BEQ
{} \bigl[ Y_n^{(1)}, Y_m^{(2)} \bigr] \:=\:
\delta_{n+m,0}\, \bigl( 3\delta_{n,0} -2 \bigr)\, \Theta, \quad
n,m\in\{\pm 1,0\}, 
\EEQ 
with a central generator $\Theta$. This is called the {\it exotic Galilean
conformal algebra} $\mbox{\sc ecga} = \mbox{\sc cga}(2) + \mathbb{C} \Theta$ in the physics literature. 
A representation as time-space transformation of {\sc ecga} is, 
with $n\in\{\pm 1,0\}$ and $j,k\in\{1,2\}$ \cite{Martelli09,Cherniha10,Henkel14} 
\BEA
{} X_n &=& -t^{n+1}\partial_t - (n+1) t^n \vec{r}\cdot\vec{\nabla}_{\vec{r}} -
x(n+1) t^n  - (n+1)n t^{n-1} \vec{\gamma}\cdot\vec{r} - (n+1)n
\vec{h}\cdot\vec{r}
\nonumber \\
Y_n^{(j)} &=& - t^{n+1}\partial_j -(n+1) t^n \gamma_j - (n+1) t^n h_j -n(n+1)\theta \eps_{jk} r_k
 \\
R_0^{(12)} &=& - \bigl( r_1 \partial_{2} -  r_2 \partial_{1} \bigr)
- \bigl( \gamma_1 \partial_{\gamma_2}  -
\gamma_2\partial_{\gamma_1}\bigr) - \frac{1}{2\theta} \vec{h}\cdot
\vec{h} \nonumber
\EEA
The components of the vector $\vec{h}=(h_1,h_2)$ satisfy $[h_i,hj]=\eps_{ij}\theta$, 
where $\theta$ is a constant, $\eps$ is the totally antisymmetric $2\times 2$ tensor and 
$\eps_{12}=1$.\footnote{An infinite-dimensional extension of 
{\sc ecga} does not appear to be possible.} The dynamical exponent $z=1$. 
Because of Schur's lemma, the central generator $\Theta$ can be
replaced by its eigenvalue $\theta\ne 0$. The \mbox{\sc ecga}-invariant Schr\"odinger operator\index{Schr\"odinger operator} is
\BEQ 
{\cal S} = -\theta X_{-1} + \eps_{ij} Y_0^{(i)} Y_{-1}^{(j)} 
= \theta \partial_t + \eps_{ij} \left( \gamma_i + h_i \right) \partial_j
\EEQ
with $x=x_{\phi}=1$. The requirement that these representations should be unitary gives the bound $x\geq 1$ \cite{Martelli09}. 
Co-variant $n$-point functions and their applications have been studied in great detail. 

{\bf 4.} The common sub-algebra of $\mathfrak{sch}(d)$ and $\mbox{\sc cga}(d)$ 
is called the {\em ageing algebra}\index{ageing algebra}
$\mathfrak{age}(d):=\langle X_{0,1},Y_{\pm\demi}^{(j)},M_0,R_0^{(jk)}\rangle$ with $j,k=1,\ldots,d$ 
and does {\em not} include time-translations. 
Starting from the representation (\ref{5}), only the
generators $X_n$ assume a more general form \cite{Picone04,Henkel06a}  
\BEQ \label{18}
X_n  = - t^{n+1}\partial_t - \frac{n+1}{2} t^n \vec{r}\cdot\vec{\nabla}_{\vec{r}} - \frac{n+1}{2} x t^n 
- n(n+1)\xi t^{n} -\frac{n(n+1)}{4}{\cal M} t^{n-1} \vec{r}^2 
\EEQ
such that $z=2$ is kept from (\ref{5}). 
The invariant Schr\"odinger operator now becomes 
${\cal S}=2{\cal M}\partial_t - \partial_r^2 +2{\cal M}t^{-1}\left(x+\xi-\demi\right)$, 
but without any constraint, neither on $x$ nor on $\xi$ \cite{Stoimenov13}. 
Co-variant $n$-point functions can be derived as before \cite{Henkel06a,Henkel10,Minic12}, but
we shall include these results with those to be derived from more general representations 
in the next sections. The absence of time-translations
is particular appealing for application to dynamical critical phenomena, such as physical ageing, 
in non-stationary states far from equilibrium, see \cite{Henkel10}. 

In figure~\ref{fig1}, the root diagrammes \cite{Knapp86} of the Lie algebra (a) $\mathfrak{age}(1)$, 
(b) $\mathfrak{sch}(1)$ and (c) $\mbox{\sc cga}(1)$ are shown, 
where the generators (roots) are represented by the black dots. 
This visually illustrates that the Schr\"odinger and conformal galilean algebras are {\it not} isomorphic, 
$\mathfrak{sch}(d)\not\cong\mbox{\sc cga}(d)$. 

Comparing figure~\ref{fig1}a with figure~\ref{fig1}c, a different representation of $\mbox{\sc cga}(1)$ can be identified. 
This representation is spanned by the generators $X_{0,1}, Y_{\pm 1/2},M_0$ from 
(\ref{5}), along with a new generator $V_+$, and leads to
a dynamic exponent $z=2$ \cite{Henkel03a}. It is not possible to extend 
this to a representation of $\mathfrak{av}(1)$ \cite{Henkel06b}. 
Explicit expressions of $V_+$ will be given in section~4. 

These algebras also appear in more systematic approaches, 
either from a classification of non-relativistic limits of conformal symmetries \cite{Duval09}
or else from an attempt to construct all possible infinitesimal local scale transformations \cite{Henkel97,Henkel02}. 

{\bf 5.} In non-equilibrium statistical mechanics \cite{Taeuber14}, 
one considers often equations under the form of a stochastic Langevin equation, viz. (we use the so-called `model-A' dynamics with 
a non-conserved order-parameter)
\BEQ
2{\cal M}\partial_t \phi = \vec{\nabla}_{\vec{r}}\cdot\vec{\nabla}_{\vec{r}}\phi - \frac{\delta {\cal V}[\phi]}{\delta \phi} +\eta
\EEQ
for a physical field $\phi$ (called the {\em order parameter}), and where $\delta/\delta\phi$ stands for a functional derivative. 
Herein, ${\cal V}[\phi]$ is the Ginzburg-Landau potential and $\eta$ is a white noise, i.e.
its formal time-integral is a brownian motion. 
In the context of Janssen-de Dominicis theory, see \cite{Taeuber14}, 
this can be recast as the variational equation of motion of the functional
\BEA
{\cal J}[\phi,\wit{\phi}] &=& {\cal J}_0[\phi,\wit{\phi}] + {\cal J}_b[\wit{\phi}] \nonumber \\
{\cal J}_0 [\phi,\wit{\phi}] &=& \int_{\mathbb{R}_+\times\mathbb{R}^d}\!\!\!\D t\D\vec{r}\: \wit{\phi}\left(2{\cal M}\partial_t \phi
- \vec{\nabla}_{\vec{r}}\cdot\vec{\nabla}_{\vec{r}} \phi +\frac{\delta {\cal V}[\phi]}{\delta\phi} \right)
 \\
{\cal J}_b [\wit{\phi}] &=& - T\int_{\mathbb{R}_+\times\mathbb{R}^d}\!\!\!\D t \D\vec{r}\: \wit{\phi}^2(t,\vec{r}) 
-\demi\int_{\mathbb{R}^{2d}}\!\D\vec{r}\D\vec{r}'\: \wit{\phi}(0,\vec{r})c(\vec{r}-\vec{r}')\wit{\phi}(0,\vec{r}')
\nonumber
\EEA
where the term ${\cal J}_0[\phi,\wit{\phi}]$ contains the deterministic terms 
coming from the Langevin equation and ${\cal J}_b [\wit{\phi}]$
contains the stochastic terms generated by averaging over the thermal noise and the 
initial condition, characterised by an initial correlator
$c(\vec{r})$.\footnote{Although it might appear that $z=2$, 
the renormalisation of the interactions, required in interacting field-theories,
can change this and produce non-trivial values of $z$, see e.g. \cite{Taeuber14}.}  
In particular, by adding an external source term $h(t,\vec{r})\phi(t,\vec{r})$ 
to the potential ${\cal V}[\phi]$, one can write the
two-time linear response function as follows (spatial arguments are suppressed for brevity) 
\BEQ \label{19}
R(t,s) = \left.\frac{\delta\langle\phi(t)\rangle}{\delta h(s)}\right|_{h=0} 
= \int\!{\cal D}\phi{\cal D}\wit{\phi}\, \phi(t)\wit{\phi}(s) e^{-{\cal J}[\phi,\wit{\phi}]} 
= \left\langle \phi(t)\wit{\phi}(s)\right\rangle
\EEQ
with an explicit expression of the average $\langle .\rangle$ as a functional integral.\\[0.02truecm] 

\noindent {\bf Theorem 1.} {\rm \cite{Picone04}} 
{\it If in the functional ${\cal J}[\phi,\wit{\phi}] = {\cal J}_0[\phi,\wit{\phi}] + {\cal J}_b[\wit{\phi}]$,  
the part ${\cal J}_0$ is Galilei-invariant with non-vanishing masses and 
${\cal J}_b[\wit{\phi}]$ does not contain the field $\phi$, then
the computation of all responses and correlators can be reduced to averages which only 
involve the Galilei-invariant part ${\cal J}_0$.}\\[0.02truecm] 

\noindent {\bf Proof:} We illustrate the main idea for the calculation of the two-time response. Define the average 
$\langle  X\rangle_0 = \int\!{\cal D}\phi{\cal D}\wit{\phi}\: X[\phi] e^{-{\cal J}_0[\phi,\wit{\phi}]}$ 
with respect to the functional ${\cal J}_0[\phi,\wit{\phi}]$. Then, from (\ref{19})
\BD
R(t,s) = \left\langle \phi(t)\wit{\phi}(s) e^{-{\cal J}_b[\wit{\phi}]}\right\rangle_0 
= \sum_{k=0}^{\infty} \frac{(-1)^k}{k!} \left\langle \phi(t)\wit{\phi}(s){\cal J}_b[\wit{\phi}]^k\right\rangle_0 
= \left\langle \phi(t)\wit{\phi}(s) \right\rangle_0 
\ED
since the Bargman superselection rule (\ref{9}) implies that only the term with $k=0$ remains. 
Hence the response function $R(t,s)=R_0(t,s)$ is reduced to the expression obtained from the
deterministic part ${\cal J}_0$ of the action. 

Analogous reduction formul{\ae} can be derived for all Galilei-covariant 
$n$-point responses and correlators \cite{Picone04,Henkel10}. 
\hfill ~ q.e.d. 

This means that one may study the deterministic, noiseless truncation of the Langevin equation and its symmetries, 
provided that spatial translational- and Galilei-invariance are included therein, in order to obtain the form
of the stochastic two-time response functions, as it will be obtained from models, simulations or experiments. 

This work is organised as follows. In section~2, 
we review several distinct representations of the Schr\"odinger and conformal Galilean algebras,
discuss the associated invariant Schr\"odinger operators an co-covariant two-point functions. 
Applications to non-equilibrium statistical
mechanics and the non-relativistic AdS/CFT correspondence will be indicated. In section~3, 
the dual representation and the extensions to parabolic sub-algebras will be reviewed. 
In section~4, it will be shown how to use these, to algebraically
derive causality and long-distance properties of co-variant two-point functions. 
Conclusions are given in section~5. 

\section{Representations}

We now list several results relevant for the extension of the representations discussed in the introduction. 
The basic new fact, first observed in \cite{Minic12}, is compactly stated as follows.\\[0.02truecm] 

\noindent {\bf Proposition 1.} {\it Let $\gamma$ be a constant and $g(z)$ a non-constant function. Then the generators}
\BEQ \label{confrep}
\ell_n = - z^{n+1}\partial_z -n\gamma z^n - g(z)z^n
\EEQ
{\it obey the conformal algebra $[\ell_n,\ell_m]=(n-m)\ell_{n+m}$ for all $n,m\in\mathbb{Z}$.}\\[0.02truecm] 

The commutator is readily checked. We point out that the {\em rapidity} $\gamma$ serves as a second scaling dimension and the choice
of the function $g(z)$ can be helpful to include effects of corrections to scaling into the generators of time-space transformations. 
Next, we give an example on how these terms in the generators $\ell_n$ appear in the two-point function, co-variant under the maximal
finite-dimensional sub-algebra $\langle \ell_{\pm 1,0}\rangle$.\\[0.02truecm]

\noindent {\bf Proposition 2.} {\it If $\phi(z)$ is a quasi-primary scaling operator under the representation 
(\ref{confrep}) of the conformal algebra 
$\langle \ell_{\pm 1,0}\rangle$, its co-variant two-point function is given by, where $\vph_0$ is a normalisation constant}  
\BEQ
\left\langle \phi_1(z_1) \phi_2(z_2) \right\rangle = \vph_0\, \delta_{\gamma_1,\gamma_2}\: (z_1-z_2)^{-\gamma_1-\gamma_2} \Gamma_1(z_1) \Gamma_2(z_2), \quad
\Gamma_i(z) := z^{\gamma_i} \exp\left( -\int_1^z \!\D\zeta\: \frac{g(\zeta)}{\zeta} \right)
\EEQ

\noindent {\bf Proof:} For brevity, denote $F(z_1,z_2)=\left\langle \phi_1(z_1) \phi_2(z_2) \right\rangle$. 
Then the co-variance of $F$ is expressed by 
the three Ward identities, with $\partial_i := \partial/\partial z_i$
\BEA
\ell_{-1} F &=& \left( -\partial_1 - \partial_2 +\gamma_1 z_1^{-1} +\gamma_2 z_2^{-1} - g(z_1)  z_1^{-1} - g(z_2) z_2^{-1} \right) F \:=\: 0
\nonumber \\
\ell_{0} F &=& \left( -z_1 \partial_1 - z_2 \partial_2 - g(z_1) - g(z_2) \right) F \hspace{3.54truecm}\:=\: 0
\nonumber \\
\ell_{1} F &=& \left( -z_1^2\partial_1 - z_2^2\partial_2 -\gamma_1 z_1 -\gamma_2 z_2 - g(z_1) z_1 - g(z_2) z_2 \right) F \hspace{0.21truecm}\:=\:0
\nonumber 
\EEA
Rewrite the correlator as $F(z_1,z_2) = \Gamma_1(z_1) \Gamma_2(z_2) \Psi(z_1,z_2)$. Then the function $\Psi(z_1,z_2)$ satisfies
\BEA
\left( -\partial_1 - \partial_2 \right) \Psi &=& 0
\nonumber \\
\left( -z_1 \partial_1 - z_2 \partial_2 - \gamma_1 - \gamma_2 \right) \Psi &=& 0
\nonumber \\
\left( -z_1^2\partial_1 - z_2^2\partial_2 -2\gamma_1 z_1 -2\gamma_2 z_2 \right) \Psi &=& 0
\nonumber 
\EEA
which are the standard Ward identities of the representation (\ref{2}) of conformal invariance, 
where the $\gamma_i$ take the r\^ole of the
conformal weights. The resulting function $\Psi$ is well-known \cite{Polyakov70}. \hfill ~ q.e.d.

One can now generalise the representation (\ref{5}) of the Schr\"odinger-Virasoro algebra $\mathfrak{sv}(d)$.\\[0.02truecm]
 
\noindent {\bf Proposition 3.} {\it If one replaces in the representation (\ref{5}) the generator $X_n$ as follows}
\BEQ \label{24}
X_n = -t^{n+1}\partial_t -\frac{n+1}{2}t^n\vec{r}\cdot\vec{\nabla}_{\vec{r}} -\frac{n+1}{2}x t^n 
-n(n+1) \xi t^n - \Xi(t) t^n -\frac{n(n+1)}{4}{\cal M}t^{n-1}
\EEQ
{\it where $x,\xi$ are constants and $\Xi(t)$ is an arbitrary (non-constant) function, 
then the commutators (\ref{4}) of the Lie algebra $\mathfrak{sv}(d)$ are still satisfied.}\\[0.02truecm] 

This result was first obtained, for the maximal finite-dimensional sub-algebra $\mathfrak{sch}(d)$, by 
Minic, Vaman and Wu \cite{Minic12}, who also
further take the dependence on the mass $\cal M$ into account and write down terms of order 
${\rm O}(1/{\cal M})$ and ${\rm O}(1)$ in $v(t)$ explicitly. 
We extend this observation to $\mathfrak{sv}(d)$, but do not trace the dependence in $\cal M$ explicitly, 
although one could re-introduce it, if required.  The
proof is immediate, since all modifications of the generator $X_n$ 
merely depend on the time $t$ and none of the other generators of $\mathfrak{sv}(d)$ changes $t$. 
For the sub-algebra $\mathfrak{age}(d)\subset\mathfrak{sch}(d)$, 
the representation (\ref{18}) is a special case, with arbitrary $\xi$, but with $\Xi(t)=0$. 

It is obvious that similar extensions of the representations of time-space transformation of the other algebras, especially 
$\mathfrak{av}(d)$, its finite-dimensional sub-algebra $\mbox{\sc cga}(d)$ or the exotic algebra $\mbox{\sc ecga}$ apply. 
I do not believe it necessary to write them down explicitly here. Explicit forms of two- and three-point functions, co-variant
under either $\mathfrak{sch}(d)$ or $\mathfrak{age}(d)$, are derived in \cite{Minic12},  
where their important differences are discussed in detail. \\[0.01truecm] 

\noindent {\bf Proposition 4.} {\it Consider the representation (\ref{5}), but with the generators $X_n$ replaced by (\ref{24}), 
of the ageing algebra $\mathfrak{age}(d)$ and the Schr\"odinger algebra $\mathfrak{sch}(d)$.
The invariant Schr\"odinger operator has the form} 
\BEQ \label{25}
{\cal S} = 2{\cal M}\partial_t -\vec{\nabla}_{\vec{r}}^2 +2{\cal M}v(t),\quad v(t)=\frac{x+\xi-d/2}{t} + \frac{\Xi(t)}{t}
\EEQ
{\it such that a solution of ${\cal S}\phi=0$ is mapped onto another solution of the same equation. 
For the algebra $\mathfrak{age}(d)$, there is no restriction, neither on $x$, nor on $\xi$, nor on $\Xi(t)$. 
For the algebra $\mathfrak{sch}(d)$,
one has the additional condition $x=\frac{d}{2}-2\xi$.}

\noindent {\bf Proof:} To shorten the calculations, we restrict here to $d=1$. 
It is enough to restrict attention to the generators $X_{\pm 1,0}$, and we must reproduce
eq.~(\ref{8}) in this more general setting. We first look at $\mathfrak{age}(1)$. 
Consideration of $X_0$ gives $t\dot{v}(t) + v - \dot{\Xi}(t)=0$ and considering $X_1$ gives
$x+\xi-\demi+\Xi(t)+t\dot{\Xi}(t) -2t v(t) -t^2\dot{v}(t)=0$, where the dot denotes the derivative with respect to $t$. 
The second relation can be simplified to $x+\xi-\demi+\Xi(t) - t v(t)=0$ which gives the assertion. 
Going over to $\mathfrak{sch}(1)$, the condition
$[{\cal S},X_{-1}]=0$ leads to $\xi /t^{2} +\dot{\Xi}(t)/t - \Xi(t)/t^2 -\dot{v}(t)=0$. 
This is only compatible with the result found before for
$\mathfrak{age}(1)$, if $\xi=-x-\xi+\demi$. Hence,  we have $x=\demi-2\xi$, as claimed. \hfill ~ q.e.d.\\

\noindent {\bf Example 1:} For a physical illustration of the meaning of the explicitly time-dependent terms in the Schr\"odinger operator 
(\ref{25}), we consider the growth of an interface \cite{Barabasi95}. 
One may imagine that an interface can be created by randomly depositing particle onto a substrate. 
The height of this interface will be described by a function $h(t,\vec{r})$. One usually works 
in a co-moving coordinate system such that the average
height $\langle h(t,\vec{r})\rangle=0$ which we shall assume from now on. 
Then physically interesting quantities are either the interface width
$w(t) = \langle h(t,\vec{r})^2\rangle \sim t^\beta$, which for sufficiently long times 
$t$ defines the {\em growth exponent} $\beta$, or else
two-time height-height correlators $C(t,s;\vec{r})=\langle h(t,\vec{r}) h(s,\vec{0})\rangle$ or two-time response functions
$R(t,s;\vec{r})=\left.\frac{\delta\langle h(t,\vec{r})\rangle}{\delta j(s,\vec{0})}\right|_{j=0}$, 
with respect to an external deposition rate
$j(t,\vec{r})$. Their scaling behaviour is described by several non-equilibrium exponents \cite{Henkel10,Taeuber14}. 
Herein, spatial translation-invariance was assumed for the sake of simplicity of the notation. 

Physicists have identified several {\em universality classes} of interface growth, see e.g. \cite{Barabasi95,Taeuber14}. For the 
{\em Edwards-Wilkinson} universality class, $h$ is simply assumed 
to be a continuous function in space. Its equation of motion for the
height is just a free Schr\"odinger equation with an addition white noise. A distinct universality class is given by the celebrate 
{\em Kardar-Parisi-Zhang equation} which contains an additional term, quadratic in $\vec{\nabla}_{\vec{r}}h$. 
A lattice realisation may be obtained by requiring that the heights only take integer values such that the height difference 
on two neighbouring sites, such that $|\vec{r}_1-\vec{r}_2|=a$
where $a$ is the lattice constant, is restricted to $h(t,\vec{r}_1)-h(t,\vec{r}_2)=\pm 1$. 
An intermediate universality class is the one of
the {\em Arcetri model}, where the strong restriction of the Kardar-Parisi-Zhang model 
is relaxed in that $h$ is taken to be a real-valued function,
but subject to the constraint that the sum of its {\em slopes} 
$\sum_{\vec{r}} \langle \vec{\nabla}_{\vec{r}}h(t,\vec{r})^2 \rangle\stackrel{!}{=} {\cal N}$,  where
$\cal N$ is the number of sites of the lattice \cite{Henkel15b} 
(this is just one of the many conditions automatically satisfied in lattice 
realisations of the Kardar-Parisi-Zhang universality class). Schematically, in the continuum limit, the slopes
$u_a(t,\vec{r})=\partial h(t,\vec{r})/\partial r_a$ in the Arcetri model satisfy a Langevin equation
\BEQ \label{arcetri}
\partial_t u_a(t,\vec{r}) =  \Delta_{\vec{r}} u_a(t,\vec{r}) 
+ \mathfrak{z}(t) u_a(t,\vec{r}) + \frac{\partial}{\partial r_a}\eta(t,\vec{r})
\EEQ
$\Delta_{\vec{r}}$ is the  spatial Laplacian and $\eta$ is a white noise. 
The constraint on the slopes can be cast into a simple form by defining 
\BEQ
g(t) = \exp\left( - 2\int_0^t \!\D\tau\: \mathfrak{z}(\tau) \right) 
\EEQ
which can be shown to obey a Volterra integral equation
\BEQ \label{Volt}
d g(t) = 2 f(t) + 2 T \int_0^{t} \!\D\tau\: f(t-\tau) g(\tau) ,\quad
f(t) = d\frac{e^{-4 t} I_1(4 t)}{4 t} \left( e^{-4 t} I_0(4 t)\right)^{d-1}
\EEQ
where $T$ is the `temperature' defined by the second moment of the white noise and the $I_n$ are modified Bessel functions. 
This model is exactly soluble \cite{Henkel15b} and  the exponents of the 
interface growth are distinct from both the Edwards-Wilkinson (if $d\ne 2$) 
and the Kardar-Parisi-Zhang universality classes. 

It turns out that for all dimensions $d>0$, there is a `critical temperature' $T_c(d)>0$ such that for $T\leq T_c(d)$, 
long-range correlations build up. For example, $T_c(1)=2$ and $T_c(2)=2\pi/(\pi-2)$. 
For $T\leq T_c(d)$, the long-time solution of (\ref{Volt}) becomes $g(t)\sim t^{-\digamma}$ as $t\to\infty$. 
This is compatible with the large-time behaviour $\mathfrak{z}(t)\sim t^{-1}$ of the Lagrange multiplier in (\ref{arcetri}). 

Hence, recalling theorem~1, it is enough to concentrate on the deterministic part. This is given by the Schr\"odinger operator (\ref{25}), 
Therein, the first term in the potential $v(t)$, of order $1/t$, represents the 
asymptotic behaviour of the Arcetri model; whereas the term described by
$\Xi(t)/t$ takes into account the finite-time corrections to this leading scaling behaviour.\footnote{For $d=1$, 
the dynamics of the Arcetri model is identical \cite{Henkel15b} to
the one of the spherical Sherrington-Kirkpatrick model. The model is defined by the classical hamiltonian 
${\cal H}=-\demi\sum_{i,j=1}^{\cal N} J_{i,j} s_i s_j$, 
where the $J_{i,j}$ are  independent centred gaussian variables, of variance $\sim {\rm O}(1/{\cal N})$, 
and the $s_i$ satisfy the spherical constraint
$\sum_{i=1}^{\cal N} s_i^2 = {\cal N}$. As usual, the dynamics if given by a Langevin equation \cite{Cugl95}. 
This problem is also equivalent to the statistics of the gap to the largest eigenvalue of
a ${\cal N}\times{\cal N}$ gaussian unitary matrix \cite{Perret15,Perret15th}, for ${\cal N}\to\infty$.} \\[0.02truecm]

\noindent {\bf Example 2:} We give a different illustration of the new representations of $\mathfrak{age}(d)$ with $\xi\ne 0$ 
(and $\Xi(t)=0$). Although we shall not be able to write down explicitly the invariant Schr\"odinger operator in the form specified in
eq.~(\ref{25}), this example makes it clear that the domain of of these representations extends beyond the context of that single differential
equation.  

The physical context involved will be the {\em kinetic Ising model with Glauber dynamics}. The statistical mechanics
of the Ising model can be described in terms of discrete `spins' $\sigma_i=\pm 1$, attached to each site $i$ of a lattice. 
In one spatial dimension, one associates to each configuration $\sigma = \{ \sigma_1,\ldots,\sigma_{\cal N}\}$ of spins
an energy (hamiltonian) ${\cal H}=-\sum_{i=1}^{\cal N} \sigma_i \sigma_{i+1}$, with periodic boundary conditions $\sigma_{{\cal N}+1}=\sigma_1$. 
The dynamics of these spins is described in terms of a Markov process, such that the `time' $t\in\mathbb{N}$ is discrete. 
At each time-step, a single spin $\sigma_i$ is randomly selected
and is updated according to the Glauber rates (also referred to as `heat-bath rule') \cite{Glauber63}. 
These are specified in terms of the probabilities 
\BEQ
P\left(\sigma_i(t+1)=\pm 1\right) = \demi \left[ 1\pm \tanh\left(\frac{1}{T}\left( \sigma_{i-1}(t)+\sigma_{i+1}(t) + h_i(t) \right)\right) \right]
\EEQ
where the constant $T$ is the temperature and $h_i(t)$ is a time-dependent external field. From these probabilities alone, the
time-evolution of the average of any local observable, such as the time-dependent
202 magnetisation or magnetic correlators, can be evaluated analytically \cite{Glauber63}. In one spatial dimension, and at temperature $T=0$,
the model displays dynamical scaling and the exactly known magnetic two-time correlator and response take a simple form.  
In the scaling limit $t,s\to\infty$
with $t/s$ being kept fixed, one has \cite{Godreche00a,Lippiello00,Henkel04}
\BEA
C(t,s) &=& \left\langle \sigma_i(t) \sigma_i(s) \right\rangle \:=\: \frac{2}{\pi} \arctan \sqrt{\frac{2}{t/s-1}} 
\label{gl:Ising1C} \\
R(t,s) &=& \left. \frac{\delta \left\langle \sigma_i(t)\right\rangle}{\delta h_i(s)}\right|_{h=0} \:=\: \frac{1}{\sqrt{2\,}\,\pi}\frac{1}{\sqrt{s(t-s)}}
\label{gl:Ising1R}
\EEA
This is independent of the initical conditions (which merely enter into corrections to scaling), hence these results should be interpreted 
as being relvant to a critical point at $T_c=0$.

As a first observation, we remark that the form (\ref{gl:Ising1R}) of the auto-response function $R(t,s) =R(t, s; 0)$ 
is not compatible with the prediction eq.~(\ref{10}) of Schr\"odinger-invariance. This means that 
the representation (\ref{5}) of the Schr\"odinger-algebra $\mathfrak{sch}(1)$, with time-translation-invariance included, is
too restrictive to account for the phenomenology of the relaxational behaviour, far from a stationary
state, of the one-dimensional Glauber-Ising model.\footnote{A historical comment: we have been aware of this since the very beginning 
of our investigations, in the early 1990s.
The exact result (\ref{gl:Ising1R}) looked strange, since the time-space response of the Glauber-Ising model does have the nice form
$R(t,s;r) = R(t,s) \exp\left[ -\demi {\cal M} r^2/(t-s)\right]$, as expected from Galilei-invariance. Only several years later, we saw how
the representations of the Schrödinger algebra had to be generalised, which was only possible by giving up explicitly
time-translation-invariance \cite{Picone04,Henkel06a}.} 

In order to account for the exact results eqs.~(\ref{gl:Ising1C},\ref{gl:Ising1R}) in terms of the representation (\ref{18}) of $\mathfrak{age}(d)$, 
one first generalises the prediction (\ref{10}) of the Schr\"odinger algebra to the corresponding one of the ageing algebra as follows, 
up to normalisation \cite{Picone04}
\BEA 
R(t,s;\vec{r}) &=& \left\langle \phi(t,\vec{r}) \wit{\phi}(s),\vec{0})\right\rangle 
\:=\: R(t,s) \exp\left[-\frac{{\cal M}}{2} \frac{ \vec{r}^2}{t-s} \right]
\label{2.11} \\
&=& \delta_{x+2\xi,\wit{x}+2\wit{\xi}}\:\delta({\cal M}+\wit{\cal M})\: 
s^{-(x+\wit{x})/2}\left(\frac{t}{s}\right)^{\xi+\digamma}\left(\frac{t}{s}-1\right)^{-x-2\xi} 
\exp\left[-\frac{{\cal M}}{2} \frac{ \vec{r}^2}{t-s} \right]
\nonumber
\EEA
with $\digamma:=\demi(\wit{x}-x)+\wit{\xi}-\xi$. Herein, $\phi$ denotes the order parameter, with scaling dimensions $x,\xi$ and of mass ${\cal M}>0$.
The response field $\wit{\phi}$, with scaling dimensions $\wit{x},\wit{\xi}$ and mass $\wit{\cal M}=-{\cal M}<0$ takes over the r\^ole of the
`complex conjugate' in eq.~(\ref{10}). Spatial translation-invariance is implicitly admitted. 
Comparison of the auto-response $R(t,s)$, eq.~(\ref{2.11}), with the exact result (\ref{gl:Ising1R}) leads to the identifications
$x=\demi$, $\wit{x}=0$, $\xi=0$ and $\wit{\xi}=\frac{1}{4}$. Remarkably,
the second scaling dimension $\wit{\xi}$ of the response scaling operator $\wit{\phi}$ does not vanish –- a feature also
observed numerically in models such as directed percolation or the Kardar-Parisi-Zhang equation, see
\cite{Henkel10b,Henkel12,Henkel13} for details.
 
On the other hand, along the lines of Theorem~1, the autocorrelator at the critical point $T=T_c$ 
can be expressed as an integral of a `noiseless' three-point response, up to normalisation \cite{Henkel94}
\BEQ
C(t,s) =  \int_{\mathbb{R}_+\times\mathbb{R}^d} \!\!\!\D u \D\vec{R}\: 
\left\langle \phi(t,\vec{y}) \phi(s,\vec{y}) \wit{\phi}^{\,2}(u,\vec{R}+\vec{y})\right\rangle_{0}
\EEQ
Schr\"odinger-invariance fixes this three-point function up to a certain undetermined scaling function. 
Herein, one considers $\wit{\phi}^{\,2}$ as a new composite operator with scaling dimensions $2\wit{x}_2, 2\wit{\xi}_2$. Up to normalisation, the
autocorrelator becomes (assuming $t>s$ for definiteness) \cite{Henkel06a}
\BEA
C(t,s) &=& s^{1+d/2-x-\wit{x}_2} \left(\frac{t}{s}\right)^{x+\digamma}\left(\frac{t}{s}-1\right)^{\wit{x}_2+2\wit{\xi}_2-x-2\xi-d/2}
\nonumber \\
& & \times \int_0^1 \!\D v\: v^{2\wit{\xi}_2-\digamma} \left[ \left(\frac{t}{s}-v\right)\left(1-v^{\,}\right)\right]^{d/2-\wit{x}_2-2\wit{\xi}_2}
{\bf\Psi}\left(\frac{t/s+1-2v}{t/s-1}\right)
\nonumber \\
&=& C_0 \int_0^1 \!\D v\: v^{2\mu} \left[ \left(\frac{t}{s}-v\right)\left(1-v^{\,}\right)\right]^{2\mu} \left(\frac{t}{s}+1-2v\right)^{2\mu}
\label{2.13}
\EEA
where in the second line we recognised that the scaling function can be described in terms of the single parameter $\mu=\xi+\wit{\xi}_2$ and
there remains an undetermined scaling function ${\bf \Psi}$. Furthermore,
the autocorrelator scaling function should be non-singular as $t\to s$. This implies ${\bf \Psi}(w)\sim w^{\wit{x}_2-x-4\xi-d/2+\mu}$ for $w\gg 1$. 
The most simple case arises when this form remains valid for all $w$. Using the values of the scaling exponents identified from the autoresponse
$R(t,s)$ before, the exact $1D$ Glauber-Ising autocorrelator eq.~(\ref{gl:Ising1C}) is recovered from (\ref{2.13}), with the choice 
$\mu=-\frac{1}{4}$ and $C_0=2/\sqrt{\pi}$ \cite{Henkel06a}. 

Although the discrete nature of the Ising spins does not permit to recognise explicitly the continuum equation of motion (the underlying field theory
of the model is a free-fermion theory, and not a free-boson theory as in the first example\footnote{The specific structure
of the dynamical functional ${\cal J}[\phi,\wit{\phi}]$, see eq.~(\ref{JanssenDominicis}), of the Arcetri model 
(and, more generally, of the kinetic spherical model \cite{Picone04}) 
leads to $\xi+\wit{\xi}=0$, such that time-translation-invariance appears to
be formally satisfied, in contrast to the $1D$ Glauber-Ising model, where $\xi+\wit{\xi}=\frac{1}{4}$.}) 
this illustrates the necessity of the second scaling dimension $\xi$, of the representation (\ref{18}) of
$\mathfrak{age}(1)$. For $d\geq 2$ dimensions, there is no known analytical solution and one must turn to numerical simulations. 
The available evidence suggests that the second scaling dimensions $\xi+\wit{\xi}\ne 0$ at criticality, at least for dimensions $d<d^*=4$, 
the upper critical dimension. For details and a review of further examples, see \cite{Henkel10}. \\

\centerline{ * ~~*~~ * } ~\\[-0.8truecm]

How can the choice of the representation affect the physical interpretation~? This  
is illustrated further by considering a `lattice' representation
rather than the usually employed `continuum' representation of the Schr\"odinger algebra 
$\mathfrak{sch}(1)$. In table~\ref{tab1}, we list
the generators of the `continuum' representation (\ref{5}) along with the one of the `lattice' representation. 
Herein, the non-linear functions of
the derivative $\partial_r$ are understood to stand for their Taylor expansions. 
The origin of the name of a `lattice' representation can be understood  
when considering the generator $Y_{-1/2}$ of `spatial translations', which reads explicitly 
\BEQ
Y_{-1/2} f(t,r) = \left.\left. -\frac{1}{a}\right(f(t,r+a/2)-f(t,r-a/2)\right)  .
\EEQ
It is suggestive to interpret this as a discretized symmetric lattice derivative operator, with $a$ as a lattice constant, 
although the $X_n, Y_m$ are still generators of infinitesimal  transformations. 

\newcommand{\ru}{\rule[-2mm]{0mm}{8mm}}
\begin{table}
\begin{center}
\begin{tabular}{|c||l|l|} \hline 
\ru generator  & continuum & lattice \\ \hline 
\ru $X_{-1}$   & $- \partial_t$                            & $- \partial_t$ \\
\ru $X_0$      & $-t\partial_t - \frac{1}{2} r \partial_r$ & 
 $-t\partial_t - \frac{1}{a \cosh(\frac{a}{2} \, \partial_r )} r \sinh( \frac{a}{2} \, \partial_r)$\\
\ru $X_1$      & $-t^2\partial_t - t r \partial_r - \frac{1}{2} {\cal M}r^2$ & 
 $-t^2\partial_t-\frac{2 t}{a\cosh(\frac{a}{2}\, \partial_r)} r \sinh(\frac{a}{2} \, \partial_r)
 -\frac{{\cal M}}{2} \left(\frac{1}{\cosh(\frac{a}{2} \, \partial_r)}r\right)^2 $\\
\ru $Y_{-1/2}$ & $-\partial_r $                      & $-\frac{2}{a} \sinh( \frac{a}{2} \, \partial_r)$  \\
\ru $Y_{1/2}$  & $-t\partial_r -{\cal M} r$          & $-\frac{2 t}{a}\sinh(\frac{a}{2}\,\partial_r)-\frac{{\cal M}}{\cosh(\frac{a}{2}\,\partial_r)} r$\\
\ru $M_0$      & $-{\cal M}$                         & $-{\cal M}$ \\ \hline
\end{tabular}
\caption{The `lattice' representations of the Schr\"odinger algebra $\mathfrak{sch}(1)$, and its `continuum' representation, 
to which it reduces in the limit $a\to 0$ \cite{Henkel94b}.\label{tab1}}
\end{center}
\end{table}

The Schr\"odinger operator has the following form, in  the `lattice' representation 
\BEQ
{\cal S} =2{\cal M} \partial_t - \frac{1}{a^2} \left( e^{a\partial_r} + e^{-a\partial_r} -2 \right) .
\EEQ
and the equation ${\cal S}\phi=0$ could be viewed as a `lattice analogue' of a free Schr\"odinger equation. 

It is also of interest to write down the co-variant two-point functions. The extension of eq.~(\ref{10}) reads, 
up to a normalisation constant \cite{Henkel94b}
\BEQ \label{2.10}
\Phi(t,n) :=  \left\langle \phi_1(t_1,r_1) \phi_2^*(t_2,r_2)\right\rangle = 
\delta_{{\cal M}_1,{\cal M}_2} \delta_{x_1,x_2} \, t^{1/2 -x_1}\: e^{-t} I_n(t)
\EEQ
where $I_n$ is again a modified Bessel function, and with the abbreviations 
\BEQ
t = \frac{t_1 - t_2}{{\cal M}_1 a^2},\quad  n = \frac{r_1-r_2}{a}.
\EEQ
Herein, both $r_1$ and $r_2$ must be integer multiples of the ``lattice constant'' $a$. 
In the limit $a\to 0$, all these results reduce to 
those of the `continuum' representation, discussed in section~1. 
Again, although at first sight this looks as a physically reasonable Green's function
on an infinite chain,\footnote{See \cite{Henkel94b} for several examples of models which reproduce (\ref{2.10}).} 
the same questions as raised in relation with eq.~(\ref{10}) should be addressed. 
The extensions  discussed in the above
propositions 2-4 can be readily added, since those only concern the time-dependence of the generators. 

All representations of the Schr\"odinger algebra discussed so far have the dynamical exponent $z=2$, 
which fixes the dilatations $t\mapsto \lambda^z t$ and
$r\mapsto \lambda r$. This can be changed, however, by admitting `non-local' representations. 
We shall write them here,\footnote{In this example, one should not confuse the dynamical exponent $z=n$ with an index $n$ of the generators.} 
for the case $z=n\in\mathbb{N}$, 
in the form given for the sub-algebra $\mathfrak{age}(1)$, when the generators read \cite{Stoimenov11} 
\BEA
X_0 &=& - \frac{n}{2} t\partial_t - \demi r\partial_r - \frac{x}{2} \nonumber \\
X_1 &=& \left( - \frac{n}{2} t^2 \partial_t - tr \partial_r - (x+\xi) t \right)\partial_r^{n-2} - \frac{\cal M}{2} r^2 \nonumber \\
Y_{-1/2} &=& -\partial_r,\quad Y_{1/2} = - t \partial_r^{n-1} - {\cal M} r,\quad M_0 = -{\cal M}
\label{30}
\EEA 
and reduce to (\ref{5}) for $z=n=2$. Clearly, these generators (especially $X_1, Y_{1/2}$) 
cannot be interpreted as infinitesimal transformations
on time-space coordinates $(t,r)$ and cannot be seen as mimicking a finite transformation, 
as was still possible with the `lattice' representation
given in  table~\ref{tab1}. In \cite{Stoimenov11}, 
a possible interpretation as transformation of distribution functions of $(t,r)$ was explored, but 
the issue is not definitely settled.\\[0.02truecm] 

\noindent {\bf Proposition 5.} {\rm \cite{Stoimenov11}} 
{\it For any $n\in\mathbb{N}$, the generators (\ref{30}) of the algebra $\mathfrak{age}(d)$ 
satisfy the commutators (\ref{4}) in $d=1$ spatial dimensions, but with the only exception}
\BEQ
\left[ X_1, Y_{1/2} \right] = \frac{n-2}{2} t^2 \partial_r^{n-3} {\cal S}
\EEQ
{\it where the Schr\"odinger operator $\cal S$ is given by} 
\BEQ \label{32}
{\cal S} = n{\cal M}\partial_t - \partial_r^n +2{\cal M}\left( x+\xi +\frac{n-1}{2}\right)t^{-1}
\EEQ 
{\it These indeed generate a dynamical symmetry on the space of {solutions of the equation 
${\cal S}\phi=0$, since the only non-vanishing commutators of
$\cal S$ with the generators (\ref{30}) are}}
\BEQ
\left[ {\cal S}, X_0 \right] = -\frac{n}{2} {\cal S},\quad \left[ {\cal S}, X_1 \right] = -nt \partial_r^{n-2} {\cal S}
\EEQ

Verifying the required commutators is straightforward (but there is no known extension to a representation of $\mathfrak{agev}(1)$). 
It is possible to generalise this construction to dimensions $d>1$ and to generic dynamical
exponents $z\in\mathbb{R}_+$, but this would require the introduction of 
fractional derivatives into the generators \cite{Stoimenov13}.\footnote{See \cite{Stoimenov13} for an application to the kinetics of
the phase-separating (model-B dynamics) spherical model.} Formally, 
one can also derive the form of co-variant two-point functions 
$F(t_1,t_2;r_1,r_2)=\langle \phi_1(t_1,r_1)\phi_2^*(t_2,r_2)\rangle$.\\[0.02truecm] 

\noindent {\bf Proposition 6.} {\rm \cite{Stoimenov11}} 
{\it For $n\in\mathbb{N}$, a two-point function $F$, 
covariant under the non-local representation (\ref{30}) of the Lie algebra
$\mathfrak{age}(1)$, defined on the solution space of ${\cal S}\phi=0$, 
where ${\cal S}$ is the Schr\"odinger operator (\ref{32}), has the form
$F=\delta({\cal M}_1-{\cal M}^*_2)\,t_2^{-(x_1+x_2)/n} F(u,v,r)$, where}
\BEA
F(u,v,r) &=& (v-1)^{-\frac{2}{n}\left[(x_1+x_2)/2+\xi_1+\xi_2-n+2\right]} 
v^{-\frac{1}{n}\left[x_2-x_1+2\xi_2-n+2\right]} f\left(r u^{-1/n}\right),\quad
\mbox{\rm $n$ even} \nonumber \\
F(u,v,r) &=& (v+1)^{-\frac{2}{n}\left[(x_1+x_2)/2+\xi_1+\xi_2-n+2\right]} 
v^{-\frac{1}{n}\left[x_2-x_1+2\xi_2-n+2\right]} f\left(r u^{-1/n}\right),\quad
\mbox{\rm $n$ odd}
\EEA
{\it and the function $f(y)$ satisfies the equation $\D^{n-1} f(y)/\D y^{n-1} + {\cal M}_1 y f(y)=0$, 
and with the variables $r=r_1-r_2$, $v=t_1/t_2$ and}
\BEQ
\left\{ \begin{array}{ll} u = t_1-t_2 & \mbox{\rm ~~;~ {\it if $n$ is even}} \\ u = t_1+t_2 & \mbox{\rm ~~;~ {\it if $n$ is odd.}}
        \end{array} \right.
\EEQ

The set of admissible functions $f(y)$ will have to be restricted by imposing physically reasonable boundary conditions, 
especially $\lim_{y\to\infty} f(y)=0$. The value $z=n$ of the dynamical exponent is obvious. 

Again, one should inquire into the behaviour when $r\to\infty$. Furthermore, 
one observes that the interpretation of $u$ depends on  
whether $n$ is even or odd. In the first case, the co-variant two-point 
functions could be a physical two-time response function, 
while in the second case, it looks more like a two-time correlator, 
since it is symmetric symmetry under the exchange of the two scaling  operators. 

All representations considered here are scalar. It is possible to consider multiplets of 
scaling operators. In the case of conformal invariance,
one should formally replace the conformal weight $\Delta$ by a matrix 
\cite{Gurarie93,Saleur92,Rahimi97,Mathieu07,Mathieu07b}. 
New structures are only found if that matrix takes
a Jordan form. Analogous representations can also be considered for the Schr\"odinger 
and conformal Galilean algebras and their sub-algebras. 
Then, it becomes necessary to consider simultaneously the scaling dimensions 
$x,\xi$ and the rapidities $\vec{\gamma}$ as matrices \cite{Hosseiny10,Hosseiny10b,Hosseiny11,Moghimi00,Henkel14}. From the
Lie algebra commutators it can then be shown that these 
characteristic elements of the scaling operators are simultaneously Jordan \cite{Henkel14}. 
Several applications to non-equilibrium relaxation phenomena have been explored in the 
literature \cite{Henkel10b,Henkel12,Hyun12,Gray13}, see \cite{Henkel13} for a review. 

\section{Dual representations}

In order to understand how the causality and the large-distance behaviour of 
the co-variant two-point functions can be understood algebraically, 
it is helpful to go over to a dual description. The new dual coordinate $\zeta$ 
is related to either the scalar mass $\cal M$ for the Schr\"odinger algebra
(this fact was first noted by Giulini \cite{Giulini96} for the Galilei algebra) 
or else to the vector of the rapidities $\vec{\gamma}$ for the
conformal Galilei algebra. It will therefore be scalar or vector, respectively. The dual fields are \cite{Henkel03a,Henkel15}
\BEA
\hat{\phi}(\zeta,t,\vec{r}) &:=& \frac{1}{\sqrt{2\pi}\,} 
\int_{\mathbb{R}} \!\D{\cal M}\: e^{\II{\cal M}\zeta} \phi_{\cal M}(t,\vec{r}) 
,\quad \mbox{\rm for $\mathfrak{sch}(d)$, $\mathfrak{age}(d)$} \nonumber \\
\hat{\phi}(\vec{\zeta},t,\vec{r}) &:=& \frac{1}{(2\pi)^{d/2}} \int_{\mathbb{R}^d} \!\D\vec{\gamma}\: 
e^{\II \vec{\gamma}\cdot\vec{\zeta}} \phi_{\vec{\gamma}}(t,\vec{r}),\quad \mbox{\rm for $\mbox{\sc cga}(d)$}
\EEA

For the sake of notational simplicity, we shall almost always restrict to the one-dimensional case, 
although we shall quote some final results for a generic dimension $d$. 

{\bf 1.} From Proposition~3, the dual generators of the Schr\"odinger-Virasoro algebra take the form (with $j,k=1,\ldots,d$)
\BEA
X_n &=& \frac{\II}{4}(n+1)n\, t^{n-1} {r}^2\partial_{\zeta}
-t^{n+1}\partial_t - \frac{n+1}{2}t^n {r}\partial_r  
- \frac{n+1}{2} x t^n - n(n+1)\xi t^n -\Xi(t) t^n \nonumber \\
Y_m &=& \II \left( m + \demi\right) t^{m-1/2} r\partial_{\zeta} - t^{m+1/2} \partial_r 
\label{schMVW} \\
M_n &=&  \II t^n \partial_{\zeta}  \nonumber 
\EEA 
with $n\in\mathbb{Z}$ and $m\in\mathbb{Z}+\demi$. This acts on a $(d+2)$-dimensional space, with coordinates $\zeta,t,\vec{r}$. 
The finite-dimensional sub-algebra $\mathfrak{sch}(1)$ generates dual dynamical symmetries of the Schr\"odinger operator 
\BEQ \label{schdual}
{\cal S} = -2\II\partial_{\zeta}\partial_t - \partial_r^2 
- 2\II \left(x+\xi-\demi\right) t^{-1} \partial_{\zeta}
\EEQ
Co-variant dual three-point functions have been derived explicitly \cite{Minic12}. 

In the context of the non-relativistic AdS/CFT correspondence, 
also referred to as {\em non-relativistic holography} by string theorists, see \cite{Gray13,Dobrev14} and refs. therein, 
one rather considers a $(d+3)$-dimensional space, with coordinates $Z,\zeta,t,\vec{r}$. The time-space transforming parts of the
Schr\"odinger-Virasoro generators read (generalising Son \cite{Son08}, 
who restricted himself to the finite-dimensional sub-algebra $\mathfrak{sch}(d)$)
\BEA
X_n &=& \frac{\II}{4}(n+1)n\, t^{n-1} \left(\vec{r}^2 + Z^2\right) - t^{n+1}\partial_t 
- \frac{n+1}{2} t^n \left( \vec{r}\cdot\vec{\nabla}_{\vec{r}} + Z\partial_Z \right) \nonumber \\
Y_m^{(j)} &=& \II \left( m+\demi \right) t^{m-1/2} r_j \partial_{\zeta} - t^{m+1/2} \partial_{r_j} \label{Son} \\
M_n &=&  \II t^n \partial_{\zeta}  \nonumber \\
R_n^{(jk)} &=& - t^n \left( r_j \partial_{r_k} -  r_k \partial_{r_j} \right) \nonumber
\EEA
Clearly, the variable $Z$ distinguishes the bulk from the boundary at $Z=0$. 
Heuristically, if one replaces $Z\partial_Z \mapsto x$ and then sets
$Z=0$, one goes back from (\ref{Son}) to (\ref{schMVW}), with $\xi=0$ and $\Xi(t)=0$. 

Following Aizawa and Dobrev \cite{Aizawa10,Dobrev14}, the passage between the 
boundary and the bulk is described in terms of the eigenvalues of
the quartic Casimir operator of the Schr\"odinger algebra $\mathfrak{sch}(1)$ \cite{Perroud77}
\BEQ
C_4 = \left( 4 M_0 X_0 - Y_{-1/2}Y_{1/2} - Y_{1/2}Y_{-1/2}\right)^2 - 2 \left\{ 2M_0 X_{-1} - Y_{-1/2}^2, 2 M_0 X_1 - Y_{1/2}^2 \right\}
\EEQ
such that in the representation (\ref{5}), which lives on the boundary $Z=0$, 
one has the eigenvalue $c_4 = c_4(x) := {\cal M}^2 (2x-1)(2x-5)$. 
Since $c_4(x)=c_4(3-x)$, two scaling operators with scaling dimensions $x$ and $3-x$ 
will be related. In order to formulate the holographic
principle, which prescribes the mapping of a boundary scaling operators $\vph$ 
to a bulk scaling operator $\phi$, a necessary condition is
the eigenvalue equation (in the bulk) 
\BEQ
C_4 \phi(Z,\zeta,t,r) = c_4(x) \phi(Z,\zeta,t,r)
\EEQ
The other condition is the expected limiting behaviour when the boundary is approached 
\BEQ
\phi(Z,\zeta,t,r) \stackrel{Z\to 0}{\longrightarrow} Z^{\alpha} \vph(\zeta,t,r), \quad \alpha=x,3-x
\EEQ 

\noindent {\bf Lemma 2.} {\rm \cite{Aizawa10}} 
{\it For the Schr\"odinger algebra in $d=1$ space dimension, the holographic principle takes the form}
\BEQ
\phi(Z,\chi) = \int\!\D^3\chi'\: S_{\alpha}(Z,\chi-\chi') \vph(\chi')
\EEQ
{\it where $\chi=(\zeta,t,r)$ is a label for a three-dimensional coordinate, $\D^3\chi = \D\zeta \D t\D r$ and}
\BEQ
S_{\alpha}(Z,\chi) = \left[ \frac{4Z}{-2\zeta t+r^2}\right]^{\alpha}
\EEQ
{\it and where $\alpha=x$ or $\alpha=3-x$.}\\[0.02truecm] 

\noindent {\bf Proof:} We merely outline the main ideas. First, construct the Green's function in the bulk, by solving
\BD
\left( C_4 - c_4(x)\right) G(Z,\chi;Z',\chi') = Z'^4\, \delta(Z-Z')\delta^3(\chi-\chi')
\ED
In terms of the invariant variable
\BD
u := \frac{4 Z Z'}{(Z+Z')^2-2(\zeta-\zeta')(t-t')+(r-r')^2}
\ED
the Casimir operator becomes $C_4 = 4u^2(1-u)\partial_u^2 -8u\partial_u +5$, hence $G=G(u)$. 
Next, the ansatz $G(u)=u^{\alpha}\bar{G}(u)$
reduces the eigenvalue equation to a standard hyper-geometric equation, with solutions expressed in terms of the 
hyper-geometric function $_{2}F_{1}$. 
Finally, $S_{\alpha}(Z,\chi-\chi')=\lim_{Z'\to 0} Z'^{-\alpha} G(u)$
leads to the assertion. \hfill ~ q.e.d. 

We refer to the literature for the non-relativistic reduction and the derivation of invariant differential equations \cite{Aizawa10,Dobrev14}. 
The consequences of passing to the more general representations with 
$\xi\ne 0$ and $\Xi(t)\ne 0$ \cite{Minic12} remain to be studied. 

{\bf 2.} Starting from (\ref{schMVW}), a dual representation of the conformal Galilean algebra 
$\mbox{\sc cga}(1)$ with $z=2$ is found 
if (i) the generator $X_{-1}$ is dropped, (ii) the generator $X_1$ is
taken as follows and (iii) and adds a new generator $V_+$ \cite{Henkel03a}
\BEA
X_1 &=& {\II} {r}^2\partial_{\zeta}-t^{2}\partial_t - t {r}\partial_r - \left(x+\xi\right) t \nonumber \\
V_+ &=& -\zeta r \partial_{\zeta} - t r \partial_t 
- \left( \II\zeta t + \frac{r^2}{2} \right)\partial_r - \left( x+\xi\right) r \label{cga2rep}
\EEA
They are dynamical symmetries of the dual Schr\"odinger operator (\ref{schdual}). 

{\bf 3.} Another dual representation of the algebra $\mbox{\sc cga}(d)$ is given by (with $j,k=1,\ldots,d$)
\BEA
X_n &=& +\II(n+1)n t^{n-1} \vec{r}\cdot\partial_{\vec{\zeta}} 
-t^{n+1}\partial_t -(n+1) t^n \vec{r}\cdot \partial_{\vec{r}}  -(n+1) x t^n
\nonumber \\
Y_n^{(j)} &=& -t^{n+1}\partial_{r_j} +\II (n+1) t^n \partial_{\zeta_{j}} \label{cgaGrep} \\
R_n^{(jk)} &=& - t^n \left( r_j \partial_{r_k} -  r_k \partial_{r_j} \right) 
- t^n \left( \zeta_j \partial_{\zeta_k} -  \zeta_k \partial_{\zeta_j} \right) \nonumber
\EEA
In contrast with the representations studied so far, there are no central generators $\sim \partial_{\zeta_j}$. 

The dualisation of the `lattice' representation and the non-local representations discussed in section~2 
proceeds analogously and will not be spelt out in detail here.\\ 

\begin{figure}[tb]
\includegraphics[scale=0.30]{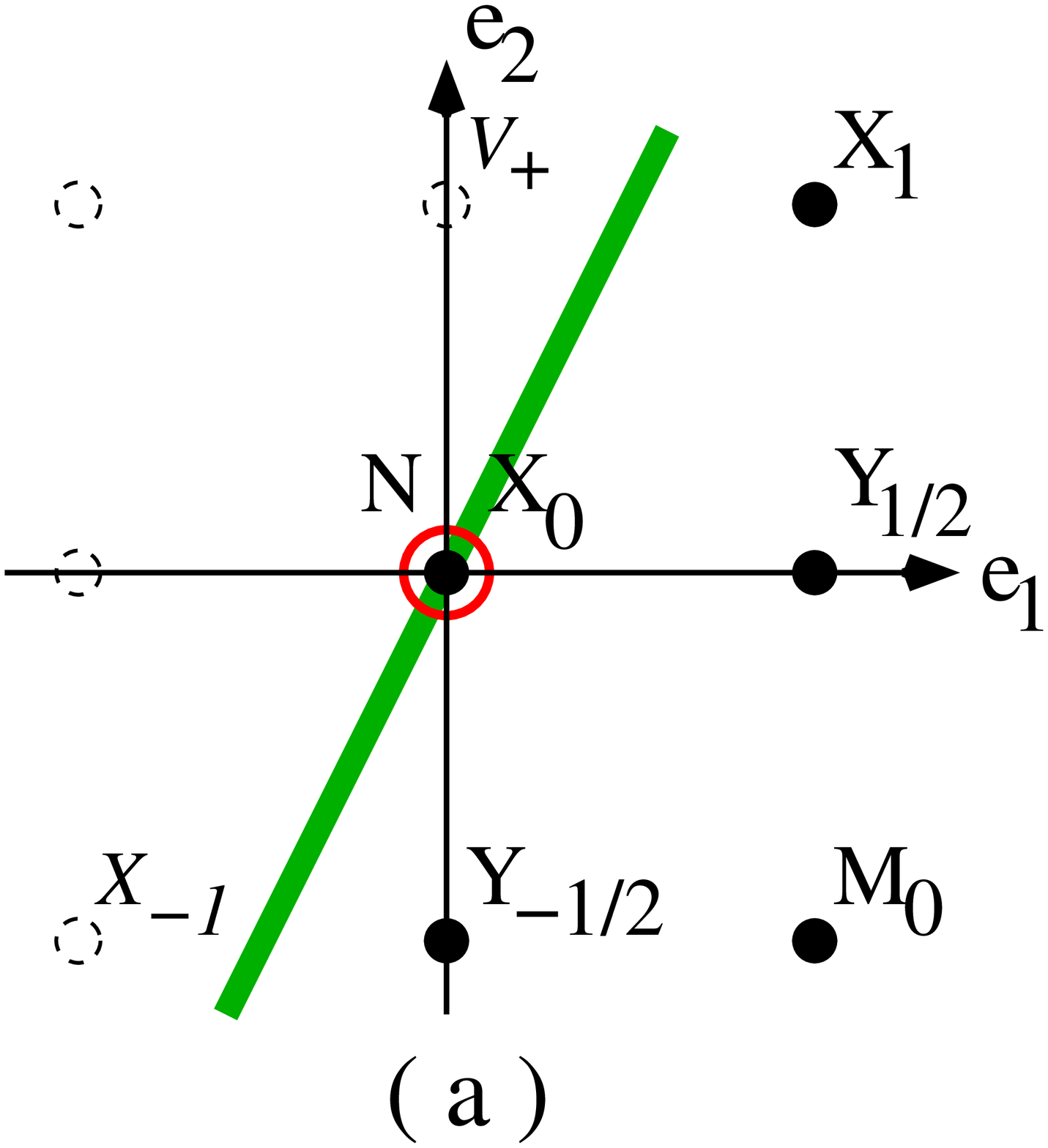}~~~\includegraphics[scale=0.30]{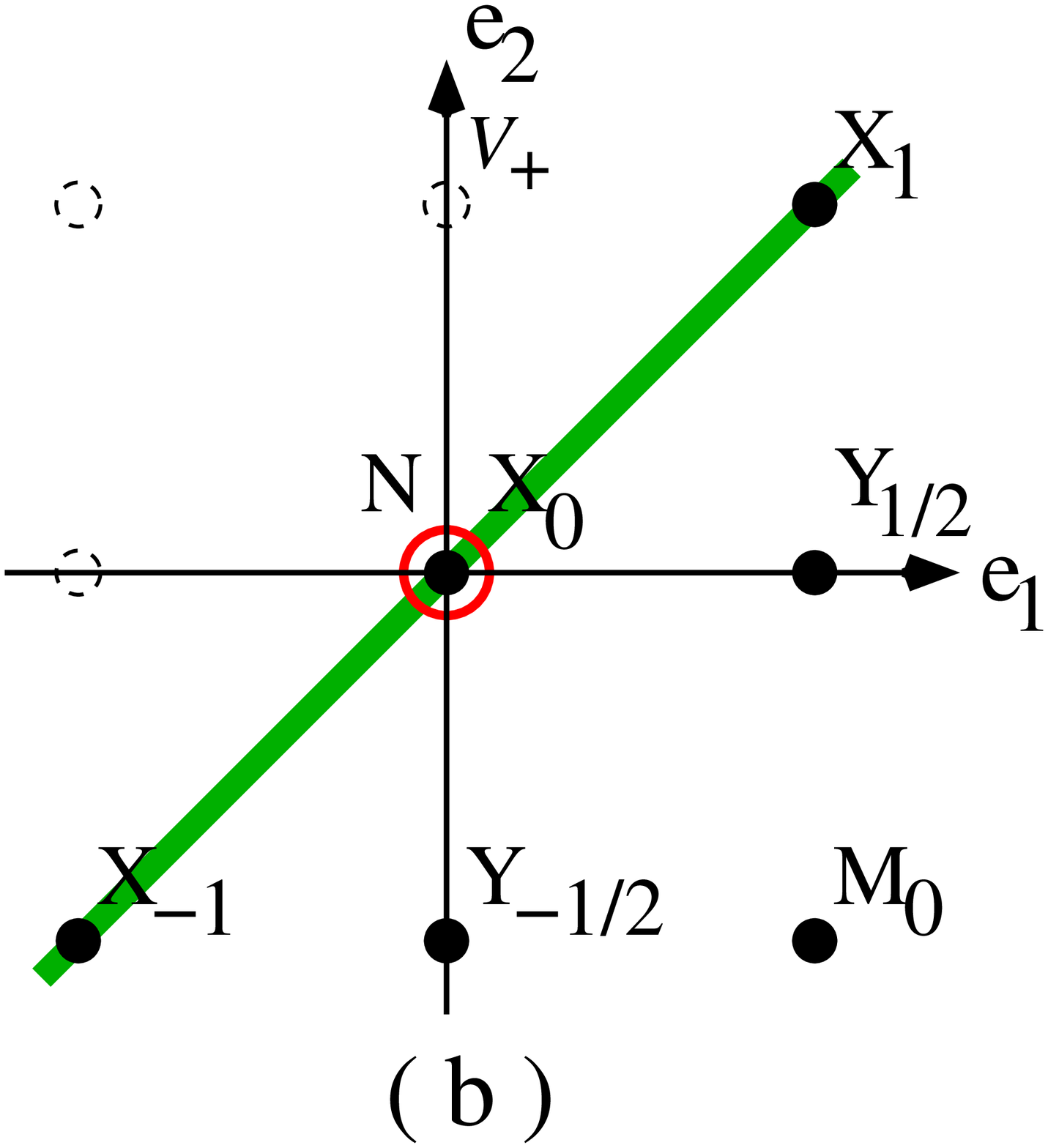}~~~\includegraphics[scale=0.30]{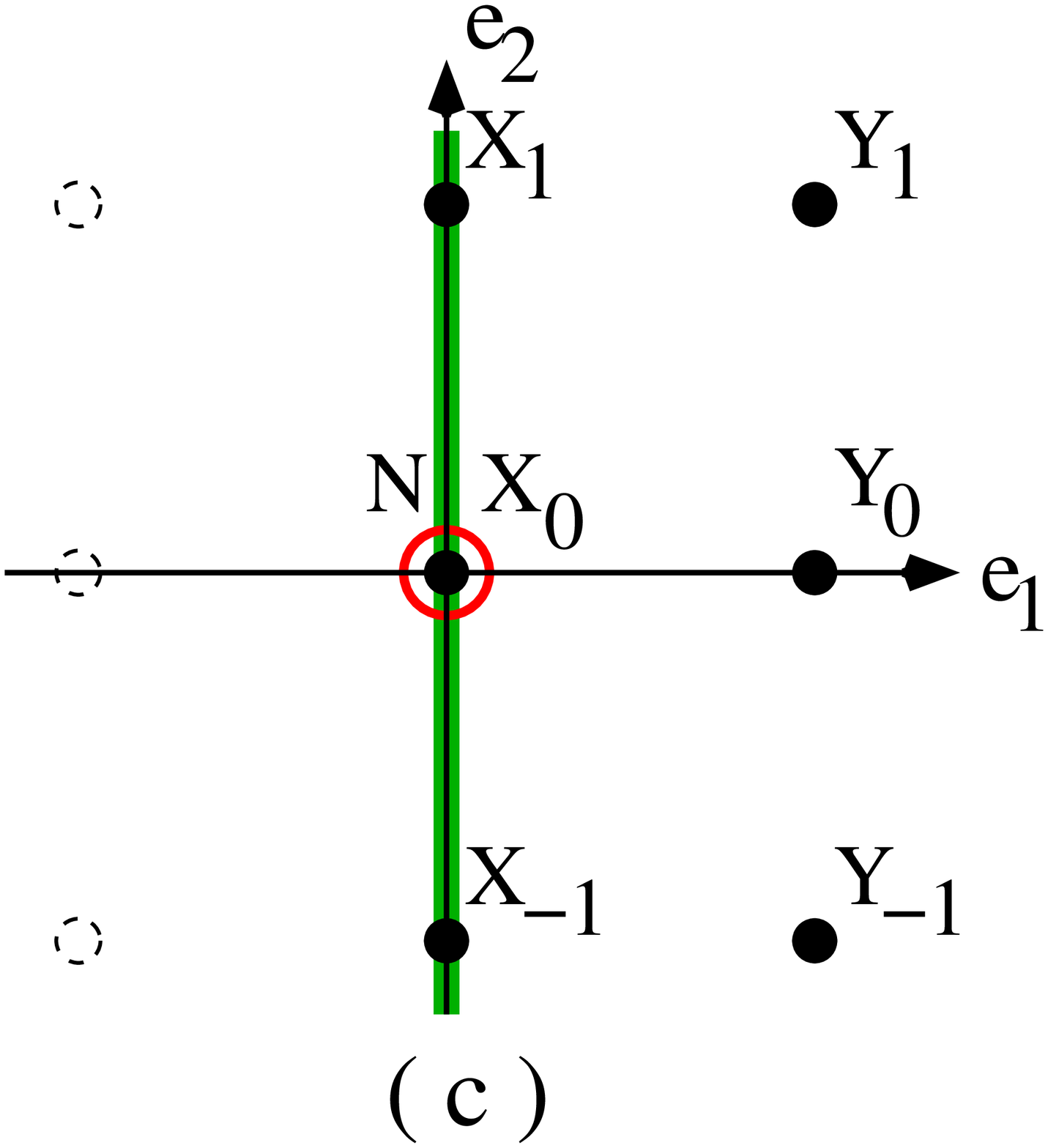} 
\caption[fig1]{Root diagrammes of the Lie algebras (a) $\mathfrak{age}(1)$, (b) $\mathfrak{sch}(1)$ and 
(c) $\mbox{\sc cga}(1)$. The generators are represented by the black filled dots. 
The red circles indicate the extra generator $N$ which extends these algebras to maximal
parabolic sub-algebras of the complex Lie algebra $B_2$. The thick green line indicates the separation 
between positive and non-positive roots. 
\label{fig1}}
\end{figure}

The other important ingredient is understood by considering the 
root diagrammes of these non-semi-simple Lie algebras, see figure~\ref{fig1}. 
Therein, it is in particular illustrated that the complexified versions 
of these algebras are all sub-algebras of the complex Lie algebra
$B_2$, in Cartan's notation \cite{Henkel03a}. In particular, it is possible to 
add further generators in the Cartan sub-algebra in order
to obtain an extension to a {\em maximal parabolic sub-algebra}. A parabolic sub-algebra is the sub-algebra of 
`positive' generators, which from a root diagramme can be identified by simply 
placing a straight line through the center (a.k.a. the 
Cartan sub-algebra). By definition, all generators which are not on the left of that line are 
called {\em positive} \cite{Knapp86}. 
In figure~\ref{fig1}, we illustrate for the three maximal parabolic sub-algebras. 
The notion of `maximal' does depend here on the precise definition of `positivity'.
For a generic slope, see figure~\ref{fig1}a, both the
generators $X_{-1}$ and $V_+$ are non-positive, and one has the maximal parabolic sub-algebra
$\wit{\mathfrak{age}}(1) = \mathfrak{age}(1) + \mathbb{C} N$. This sub-algebra is indeed maximal as a {\em parabolic} sub-algebra: for example
an extension to a Schr\"odinger algebra by including the time-translations $X_{-1}$ would no longer be parabolic, according to the
specific definition of 'positivity' used in this specific context. 
If a different definition of `positivity' is used, and the slope is now taken to be exactly unity, $X_{-1}$ is included into the positive
generators, see figure~\ref{fig1}b, and we have the maximal parabolic sub-algebra 
$\wit{\mathfrak{sch}}(1) = \mathfrak{sch}(1) + \mathbb{C} N$. Finally, and with yet a different definition of `positivity', 
where the slope is now infinite, see figure~\ref{fig1}c, one has
the maximal parabolic sub-algebra $\wit{\mbox{\sc cga}}(1) = \mbox{\sc cga}(1) + \mathbb{C}N$.
The Weyl symmetries of the root diagramme of
$B_2$ \cite{Knapp86} imply that any other maximal and non-trivial sub-algebra of $B_2$ 
is isomorphic to one of the three already given. For a formal proof, see \cite[app. C]{Henkel03a}. 

It remains to construct the operator $N$ explicitly, for each representation. 
We collect the results, coming from different sources \cite{Henkel03a,Minic12,Henkel15}.\\[0.02truecm]  

\noindent {\bf Proposition 7.} 
{\it Consider the dual representations (\ref{schMVW}) of the Schr\"odinger-Virasoro algebra, the $z=2$ dual representation
(\ref{cga2rep}) of the conformal Galilean algebra $\mbox{\sc cga}(1)$, the $z=1$ dual representation (\ref{cgaGrep}) of  
$\mbox{\sc cga}(d)$ and the dualisation 
of the non-local representation (\ref{30}) of $\mathfrak{age}(1)$, dualised with respect to $\cal M$. 
There is a generator $N$ which extends these representations to representations of the
associated maximal parabolic sub-algebra. The explicit form of the generator $N$ is as follows}
\BEQ
N = \left\{ \begin{array}{ll}
\zeta\partial_{\zeta} - t \partial_t + \xi' & \mbox{\it ~representation (\ref{schMVW}) of $\mathfrak{sch}(d)$} \\
\zeta\partial_{\zeta} - t \partial_t + \xi  & \mbox{\it ~representation (\ref{cga2rep}) of $\mbox{\sc cga}(1)$} \\
-\zeta\partial_{\zeta} - r \partial_r -\xi  & \mbox{\it ~representation of $\mbox{\sc cga}(d)$ constructed as in (\ref{cgaGrep})} \\
\zeta\partial_{\zeta} - t \partial_t + \xi' & \mbox{\it ~dualised non-local representation (\ref{30}) of $\mathfrak{age}(d)$}
\end{array} \right.
\EEQ
{\it Herein, $\xi$ is the second scaling dimension and $\xi'$ is a constant. 
These generators give dynamical symmetries of the Schr\"odinger operators $\cal S$ associated with each representation.}

\section{Causality}

It turns out that the maximal parabolic sub-algebras are the smallest Lie algebras 
which permit unambiguous statements on the causality of co-variant
two-point functions. For illustration, we shall concentrate on the dual representations (\ref{schMVW}) of $\mathfrak{sch}(d)$
and (\ref{cgaGrep}) of $\mbox{\sc cga}(d)$.\\[-0.1truecm]
 
\noindent {\bf Proposition 8.} {\rm \cite{Henkel03a,Henkel15}} 
{\it Consider the co-variant dual two-point functions. For the dual representation  (\ref{schMVW}) of $\wit{\mathfrak{sch}}(d)$, it
has the form, up to a normalisation constant}  
\BEQ \label{schN} 
\wht{F}(\zeta,t,\vec{r}) = \langle \wht{\phi}(\zeta,t,\vec{r})\wht{\phi}^*(0,0,\vec{0})\rangle = \delta_{x_1,x_2}\, |t|^{-x_1}\: 
\left( \frac{2\zeta t+\II \vec{r}^2}{|t|}\right)^{-x_1-\xi_1'-\xi_2'}
\EEQ
{\it and where translation-invariance in $\zeta,t,\vec{r}$ was used. For the dual representation (\ref{cgaGrep}) 
of $\wit{\mbox{\sc cga}}(1)$, one has, up to a normalisation constant} 
\BEQ \label{cgaN}
\wht{F}({\zeta}_+,t,{r})=\langle \wht{\phi}_1({\zeta}_1,t,{r})\wht{\phi}_2({\zeta}_2,0,{0})\rangle 
=\delta_{x_1,x_2} |t|^{-2x_1} \left({\zeta}_+ +\frac{\II {r}}{t}\right)^{-\xi_1-\xi_2}
\EEQ
{\it and where ${\zeta}_+=\demi({\zeta}_1+{\zeta}_2)$.}\\[-0.1truecm] 

This is easily verified by insertion into the respective Ward identities which express the co-variance. 
Finally, we formulate precisely
the spatial long-distance and co-variance properties of these two-point functions.\\[0.02truecm] 

\noindent {\bf Theorem 2.} {\rm \cite{Henkel03a}} 
{\it With the convention that masses  ${\cal M}\geq 0$ of scaling operators $\phi$ should be non-negative, and if
$\demi(x_1+x_2)+\xi_1'+\xi_2'>0$, the full
two-point function, co-variant under the representation (\ref{5}) of the parabolically extended Schr\"odinger algebra 
$\wit{\mathfrak{sch}}(d)$, has the form}
\BEQ \label{Thm2}
\left\langle\phi(t,\vec{r})\phi^*(0,\vec{0})\right\rangle = \delta({\cal M}-{\cal M}^*)\,\delta_{x_1,x_2}\, \Theta(t)\,t^{-x_1}\: 
\exp\left( -\frac{{\cal M}}{2}\frac{\vec{r}^2}{t}\right)  
\EEQ
{\it where the $\Theta$-function expresses the causality condition $t>0$, 
and up to a normalisation constant which depends only the mass ${\cal M}\geq 0$.}\\[-0.3truecm]  

\noindent {\bf Proof:} This follows directly from (\ref{schN}). Carrying out the inverse 
Fourier transform and using the translation-invariance in
the dual coordinate $\zeta$, one recovers the habitual two-point function multiplied 
by an integral representation of the $\Theta$-function. \hfill ~ q.e.d.

The treatment of the conformal Galilean algebra requires some further preparations, following Akhiezer \cite[ch. 11]{Akhiezer88}. 

\noindent {\bf Definition.} {\it Let $\mathbb{H}_+$ be the upper complex half-plane $w=u+\II v$ with $v>0$. 
A function $g:\mathbb{H}_+ \to \mathbb{C}$ is said to be in the {\em Hardy class $H_2^+$}, written as $g\in H_2^+$, 
if (i) $g(w)$ is holomorphic in
$\mathbb{H}_+$ and (ii) if it satisfies the bound}
\BEQ \label{3.33}
M^2 := \sup_{v>0}\: \int_{\mathbb{R}} \!\D u\: \left| g(u+\II{v})\right|^2 <\infty
\EEQ
{\it Analogously, for functions $g:\mathbb{H}_- \to \mathbb{C}$, one defines the {\em Hardy class $H_2^-$}, 
where $\mathbb{H}_-$ is the lower complex half-plane and the supremum in (\ref{3.33}) is taken over ${v}<0$.}\\[-0.1truecm]

\noindent {\bf Lemma 3.} {\rm \cite{Akhiezer88}} 
{\it If $g\in H_2^{\pm}$, then there are square-integrable functions ${\cal G}_{\pm}\in L^2(0,\infty)$ 
such that for ${v}>0$, one has the integral representation} 
\BEQ \label{Hardy}
g(w) = g(u\pm\II v) = \frac{1}{\sqrt{2\pi\,}} 
\int_0^{\infty} \!\!\D \gamma\; e^{\pm\II \gamma w}\, {\cal G}_{\pm}(\gamma)
\EEQ

We shall use eq.~(\ref{Hardy}) as follows. 
First, consider the case $d=1$. Fix $\lambda := r/t$. Now, recall (\ref{cgaN}) and write $\wht{F}=|t|^{-2x_1} \wht{f}(u)$, 
with $u=\zeta_++\II r/t$.  We shall re-write this as follows 
\BEQ
\wht{f}(\zeta_+ +\II \lambda) =: f_{\lambda}(\zeta_+)
\EEQ
and concentrate on the dependence on $\zeta_+$ (the eventual extension to $d>1$ will be obvious).\\[-0.2truecm]
 
\noindent {\bf Proposition 9.} {\rm \cite{Henkel15}} 
{\it Let $\xi := \demi(\xi_1+\xi_2) > \frac{1}{4}$. If $\lambda>0$, then $f_{\lambda}\in H_2^+$
and if $\lambda<0$, then $f_{\lambda}\in H_2^-$.} \\[-0.20truecm]

\noindent {\bf Proof:} 
The holomorphy of $f_{\lambda}$ being obvious, we merely must verify the bound (\ref{3.33}). Let $\lambda>0$. Clearly, 
$\left|f_{\lambda}(u+\II{v})\right|=\left| (u+\II( v+\lambda))^{-2\xi}\right| = \left( u^2 + ({v}+\lambda)^2 \right)^{-\xi}$. 
Hence, computing explicitly the integral,   
\BD
M^2 = \sup_{{v}>0} \int_{\mathbb{R}} \!\D u\: \left| f_{\lambda}(u+\II {v})\right|^2 
=  \frac{\sqrt{\pi\,}\: \Gamma(2\xi-\demi)}{\Gamma(2\xi)} 
\sup_{{v}>0} \left({v}+\lambda\right)^{1-4\xi}
< \infty
\ED
since the integral converges for $\xi>\frac{1}{4}$. For $\lambda<0$, the argument is similar. \hfill ~ q.e.d.

We can now formulate the second main result.
 
\noindent {\bf Theorem 3.} {\rm \cite{Henkel15}} 
{\it The full two-point function, co-variant under the representation (\ref{13}) 
of the parabolically extended conformal Galilean algebra $\wit{\mbox{\sc cga}}(d)$, 
has the form}
\BEQ \label{Thm3}
\left\langle \phi_1(t,\vec{r}) \phi_2(0,\vec{0}) \right\rangle = 
\delta_{x_1,x_2} \delta(\vec{\gamma}_1-\vec{\gamma}_2)\, 
|t|^{-2x_1} \exp\left( -2\left|\frac{\vec{\gamma}_1\cdot\vec{r}}{t}\right|\:\right) 
\EEQ
{\it up to a normalisation constant, depending only on the absolute value of the rapidity vector $\vec{\gamma}_1$.}

\noindent {\bf Proof:} Since the final result is rotation-invariant, because of the representation (\ref{13}), 
it is enough to consider the case $d=1$. Let $\lambda>0$. 
From (\ref{Hardy}) of Lemma~3 we have 
\BD
\sqrt{2\pi}\wht{f}(\zeta_+ +\II \lambda) 
= \int_0^{\infty} \!\D \gamma_+ \: 
e^{\II(\zeta_+ +\II\lambda)\gamma_+} \wht{{\cal F}_+}(\gamma_+)
= \int_{\mathbb{R}} \!\D \gamma_+ \: \Theta(\gamma_+)\, 
e^{\II(\zeta_+ +\II\lambda)\gamma_+} \wht{{\cal F}_+}(\gamma_+)
\ED
Now return from the dual two-point function $\wht{F}$ to the original one. Let $\zeta_{\pm} := \demi\left(\zeta_1\pm\zeta_2\right)$. 
We find, using also that $x_1=x_2$ 
\BEA
F &=& \frac{|t|^{-2x_1}}{\pi\sqrt{2\pi}} \int_{\mathbb{R}^2} \!\D\zeta_+ \D\zeta_-\: 
e^{-\II(\gamma_1+\gamma_2)\zeta_+}\, e^{-\II(\gamma_1-\gamma_2)\zeta_{-}} 
\int_{\mathbb{R}} \!\D\gamma_+\: 
\Theta(\gamma_+)\wht{{\cal F}_+}(\gamma_+)\, e^{-\gamma_+ \lambda}\, e^{\II\gamma_+\zeta_+}
\nonumber \\
&=& \frac{|t|^{-2x_1}}{\pi\sqrt{2\pi}} 
\int_{\mathbb{R}} \!\D\gamma_+\:  \Theta(\gamma_+)\wht{{\cal F}_+}(\gamma_+)\, e^{-\gamma_+ \lambda}
\int_{\mathbb{R}}\!\D\zeta_{-}\: e^{-\II(\gamma_1-\gamma_2)\zeta_{-}}
\int_{\mathbb{R}}\!\D\zeta_{+}\: e^{\II(\gamma_+-\gamma_1-\gamma_2)\zeta_+}
\nonumber \\
&=& \delta(\gamma_1-\gamma_2) \Theta(\gamma_1) F_{0,+}(\gamma_1)\, e^{-2\gamma_1 \lambda}\, |t|^{-2x_1} \nonumber
\EEA 
where in the last line, two $\delta$-functions were used and $F_{0,+}$ contains the unspecified dependence on 
the positive constant $\gamma_1$. An analogous argument applies for $\lambda<0$. \hfill ~ q.e.d.


\section{Conclusions}

Results on relaxation phenomena in non-equilibrium statistical physics 
and the associated dynamical symmetries have been reviewed. 
By analogy with conformal invariance which applies to equilibrium critical phenomena, 
it is tempting to try to extend the generically
satisfied dynamical scaling to a larger set of dynamical symmetries. If this is possible, 
one should obtain a set of co-variance conditions, to
be satisfied by physically relevant physical $n$-point functions. In contrast to equilibrium critical phenomena, it turned out that in 
non-equilibrium systems, each scaling operator must be characterised {\em at least} in terms of {\em two} independent scaling dimensions. 

A straightforward realisation of this programme  
leads to difficulties for a consistent physical interpretation, related to
to the requirement of a physically sensible large-distance behaviour. 
Attempting to write down  Ward identities
for the $n$-point functions, one implicitly assumes that these depend holomorphically 
on their time-space arguments, e.g. \cite{Hille76}. 
However, the constraint of causality, required for a reasonable two-time response 
function $R(t_1,t_2)$, renders $R(t_1,t_2)$ non-holomorphic in the time difference $t_1-t_2$. 
As a  possible solution of this difficulty, we propose to go over to 
dual representations with respect to either the `masses' or the
`rapidities', which are physically dimensionful parameters of the 
representations of the dynamical symmetry algebras considered. If furthermore
the dynamical symmetry algebras can be extended to a maximal parabolic 
sub-algebra of a semi-simple complex Lie algebra, then causality conditions can be derived, 
which also guarantee the requested fall-off at large distances. 

This suggests that the {\em dual} scaling operators, rather than the original ones,  
might posses interesting holomorphic properties which should be further explored. This observation might also become of
interest in further studies of the holographic principle. 

Specifically, we considered representations of (i) the Schr\"odinger algebra $\wit{\mathfrak{sch}}(d)$, 
where the co-variant two-point functions (\ref{Thm2})
have the causality properties of two-time linear {\em response functions} 
and also representations of (ii) the conformal Galilean algebra
$\wit{\mbox{\sc cga}}(d)$, where the two-point functions (\ref{Thm3}) 
have the symmetry properties of a two-time {\em correlator}. 

Although this has not yet been done explicitly, we expect that 
the techniques reviewed here can be readily extended to several physically 
distinct representations of these algebras, see e.g. \cite{Ivash97,Hartong14,Setare12,Stoimenov15} 
for examples.\footnote{How should one dualise in the {\sc ecga}~? With respect to $\theta$ or to the rapidity vector $\vec{\gamma}$~?}


\noindent {\bf\large Acknowledgements.}
I warmly thank R. Cherniha, X. Durang, A. Hosseiny, S. Rouhani, S. Stoimenov and J. Unterberger 
for numerous discussions which helped me to 
slowly arrive at the synthesis presented here. 
This work was partly supported by the Coll\`ege Doctoral Nancy-Leipzig-Coventry
(Syst\`emes complexes \`a l'\'equilibre et hors \'equilibre) of UFA-DFH.

\end{document}